\documentclass[preprint,3p,11pt]{elsarticle}

\pdfoutput=1

\usepackage{geometry}
\usepackage{amssymb,amstext}
\usepackage{graphics,graphicx}

\usepackage{placeins}  %
\newcommand{\beq}{\begin{equation}}
\newcommand{\eeq}{\end{equation}}
\newcommand{\bea}{\begin{eqnarray}}
\newcommand{\eea}{\end{eqnarray}}

\newcommand{\calC}{\mathcal{C}}

\newcommand{\barm}{{\overline m}}
\newcommand{\nP}{{\hat P}}
\newcommand{\rgtt}{\mbox{rhs}(g_{tt})}
\newcommand{\eff}{\mbox{\scriptsize eff}}
\newcommand{\mys}{n}
\newcommand{\myPhi}{d}  %
\newcommand{\myPi}{k}   %

\begin{document}

\begin{frontmatter}

\title{A pseudospectral matrix method for \\ time-dependent tensor fields
  on a spherical shell}

\author{Bernd Br\"ugmann}

\address{Theoretical Physics Institute, University of Jena, 07743 Jena, Germany}

\begin{abstract}
  We construct a pseudospectral method for the solution of
  time-dependent, non-linear partial differential equations on a
  three-dimensional spherical shell.  The problem we address is the
  treatment of tensor fields on the sphere.  As a test case we
  consider the evolution of a single black hole in numerical general
  relativity.  A natural strategy would be the expansion in tensor
  spherical harmonics in spherical coordinates.  Instead, we consider
  the simpler and potentially more efficient possibility of a double
  Fourier expansion on the sphere for tensors in Cartesian
  coordinates.  As usual for the double Fourier method, we employ a
  filter to address time-step limitations and certain stability
  issues. We find that a tensor filter based on spin-weighted
  spherical harmonics is successful, while two simplified,
  non-spin-weighted filters do not lead to stable evolutions.  The
  derivatives and the filter are implemented by matrix multiplication
  for efficiency. A key technical point is the construction of a
  matrix multiplication method for the spin-weighted spherical
  harmonic filter. As example for the efficient parallelization of the
  double Fourier, spin-weighted filter method we discuss an
  implementation on a GPU, which achieves a speed-up of up to a factor
  of 20 compared to a single core CPU implementation.
\end{abstract}


\end{frontmatter}

\section{Introduction}
\label{introduction}

Spectral methods are applicable to a wide range of partial
differential equations, e.g.\ \cite{Boy01}. We consider the case of
time-dependent tensor fields on a three-dimensional spherical shell.
The field equations are assumed to be non-linear without giving rise
to shocks, hence we choose pseudospectral collocation methods. 
Non-linearity and time-dependence may necessitate the use of
filters (or some alternative) to stabilize the method
\cite{Boy01,HesGotGot07}. Furthermore, as typical for spectral
methods, the domain influences the choice of basis functions, which in
turn matters for the computation of derivatives and for the
construction of filters.

The specific application considered in this work is a test case for
numerical general relativity, a single Schwarzschild black hole. This
is a vacuum solution of the Einstein field equations, which in adapted
coordinates is spherically symmetric and static. However, when
implemented on a 3d grid with the full evolution equations, some
non-trivial time evolution including deviations from sphericity can
occur. In particular, unstable modes leading to a failure of the
evolution after a finite time can and do appear if the problem
is not formulated with due care, which makes this example a valuable test
case, e.g.\ \cite{AlcBru00}. 
Here we study the formulation given in \cite{LinSchKid05}, which
is a first order in time and space reformulation of the Einstein
equations in the generalized harmonic gauge (GHG). 
The GHG system including modifications for stability is an example
for the class of problems that can be written in the form
\beq
\partial_t u^\mu + {A^{i\mu}}_\nu(u) \partial_i u^\nu = S^\mu(u), 
\label{deltu}
\eeq 
where $u^\mu(t,x^i)$ is the vector of variables,
$\partial_i=\partial/\partial x^i$, and a summation over up/down
indices is assumed ($i=1,2,3$). The coefficient matrices
${A^{i\mu}}_\nu$ and $S^\mu$
may depend on $u^\mu$ but not on its derivatives. 
For the GHG equations, greek indices label the 50 fields
($\mu=1,\ldots,50$) representing specific tensor components of the
field degrees of freedom. Depending on how the gauge is treated, this
number increases to 54 or 58. Specifically, the GHG system involves
rank 1, 2, and 3 tensors. 
We collect the relevant details in App.~A. 
Spectral methods in numerical relativity are reviewed e.g.\ in
\cite{BonFriGou96,GraNov09}.
The computational method discussed in this work depends in part on
the form of (\ref{deltu}), on that the fields are tensor components,
and on the choice of a spherical shell as the domain.  Other details
of the physics should only be of secondary importance and not affect
the generality of the discussion.

The specific domain under consideration, spherical shells, influences
the choice of basis functions. We choose a Chebyshev basis for the
radial direction.  For the two angular directions, the standard choice
for scalar fields is spherical harmonics, leading to a ``CY''-basis on
the 3d shell.  For tensor fields, one possibility is to employ
spin-weighted spherical harmonics
on the sphere, i.e.\ a ``CYn''-basis, where ``Yn'' indicates that
spin-weighted spherical harmonics are used. A general rank $n$ tensor
(a tensor with $n$ indices) can be decomposed in a linear combination
of spin-weight $0,\pm1,\ldots,\pm n$ spherical harmonics.
For recent work on tensor expansions with a connection to
relativity, see \cite{NovCorVas09}. 

However, especially for tensor fields, other choices of basis are
possible and sometimes even advantageous. In this work we explore the
suitability of a CFF basis, where ``FF'' stands for a double Fourier
basis on the sphere \cite{Mer73,For95,SpoTaySwa98,Che00}. The double
Fourier method includes a filter to address the clustering of points 
near the poles.
The basic choice is between the ``ideal'' filter of spherical harmonic
projection and simpler, less costly methods. 
A Y-filter is the projection on a finite number of spherical harmonics 
consisting of a forward and backward spherical harmonic transform.
The CFF basis with a
Y-filter can be equivalent to the CY method
\cite{SpoTaySwa98,SwaSpo00}. 
For tensor fields, in some cases a CY-basis with a Yn-filter is a
possible solution, see \cite{LinSchKid05} for the black hole example.
For a different formulation of the black hole problem, a CFF method
with a Y-filter has been considered in \cite{Tic06,Tic09}, although with
evolutions that are not as stable as in \cite{LinSchKid05}. 

The goal and result of the present paper is a CFF method with a
Yn-filter for tensor fields on a shell. Our method results in
long-term stable evolutions for the single black hole example
comparable to \cite{LinSchKid05}, although there remains some slow,
residual linear growth that we do not study further in this work.

Part of the rationale behind the CFF basis \cite{SwaSpo00} is that
computing partial derivatives is simpler and usually more efficient
than for a CYn or CY basis. While there exist fast Legendre transforms
to implement the spherical harmonic derivatives, they involve a higher
overhead than fast Fourier transforms, in particular for small
transform sizes. However, although the CFF method avoids spherical harmonics
in the derivatives, we choose to apply
spherical harmonics as a {\em filter}. Since the FF basis on the
sphere does not have uniform areal resolution, some type of spherical
harmonic filter can be essential to alleviate the severe time-step
restrictions due to the clustering of points near the poles in the FF
basis.  Comparing a CY method to a CFF with Y-filter method, the
latter can be more efficient since the spherical harmonic transform is
only used on the fields, while for the CY method a larger number of
spherical harmonic transforms is required for the derivatives of the
fields. 
Also, it can be easier to optimize a Y-filter, or to find alternatives
to Y-filters, rather than to optimize spherical harmonic transforms
per se.

For the particular treatment of tensor fields that we consider, a
tensor Yn-filter plays one further role, in addition to projecting onto a
uniform area basis and to filtering for stability of the non-linear
field evolution. To avoid coordinate singularities, it is convenient
to express the tensor components with respect to global Cartesian
coordinates, $(x,y,z)$, while the collocation grid is based on
spherical coordinates, $(r,\theta,\phi)$. For example, the
Cartesian components of a smooth vector field are smooth at the poles
of the spherical grid, implying spectral convergence in a CFF or CY
basis, where each component of the vector is expanded as if it were a
scalar field. While spectrally convergent, 
the Cartesian components represent a mixture of different spin-weights
that is not properly handled by the scalar Y-filter.
In particular, the CY method of
\cite{LinSchKid05}
displays a long-term instability linked to the combination of the
Cartesian components with the Y-filter. This instability was noted in
a related context \cite{KidLinSch04} and cured by a tensor spherical
harmonic filter in the examples of \cite{KidLinSch04,LinSchKid05},
although details of the instability or the implementation were not
given.

The main topic of the present work is the double Fourier method
combined with a spin-weighted spherical harmonic filter for tensor
fields. Since we may need a Yn-filter for stability anyway, this paper
examines the question whether we can do away with the complications of
Y-derivatives and Yn-derivatives completely. Can we take two shortcuts
(the FF basis and Cartesian components) and clean up with one trick
(the Yn-filter) later? In the example considered, the answer is yes,
and the method realizes the efficiency and simplicity bonus of the CFF
method with Y-filter for scalar fields. To our knowledge, while there
is literature on both the CFF method and the construction of Yn-filters,
there is no description yet of a CFF method combined with a Yn-filter
for $|n|>1$.

An important aspect of the proposed CFF/Yn-filter method is its
efficient implementation.
Since we consider a collocation method in 3d, say with $N^3$ points,
we cannot handle very large $N$ anyway. With regard to computing
1d derivatives on a 3d grid, our task is a small $N$ problem, say
$N\lesssim100$, in contrast to 2d or 1d problems with much larger
$N$.  Also, in our example exponential convergence of the solution
usually means that double precision round-off error is reached for
$N\approx 40$, since there are no features on a smaller scale to be
resolved.
If there are local features to be resolved (in the black hole example,
waves of small wavelength travelling to infinity), the recommended
strategy for efficiency is not to use large $N$ on a single domain,
but rather to take a step towards ``spectral elements'' and to
decompose the domain into several nested spherical shells. 
Therefore, with domain decomposition in mind for efficient 3d methods,
one relevant test case to consider is that of a single domain 
where the number of points in each direction is comparatively small, with
$N\lesssim100$.

Given a small $N$ problem, we are led to consider matrix methods for
the computation of derivatives and filters \cite{For98,Tre00}.
The operation count of a typical implementation of the partial
differential equation (\ref{deltu}) is dominated by the computation of
the spectral derivatives. For the Chebyshev and Fourier bases, we
compute derivatives using Fourier transforms (FTs), where the standard
choice for an efficient algorithm is the fast Fourier transform (FFT).
However, it is also well-known that for sufficiently small $N$ a FT by
direct matrix multiplication can be faster than a FFT, since it avoids
a certain overhead, e.g.\ \cite{For98,Boy01}.
Furthermore, fast methods for the Legendre transform that is part of
spherical harmonic filtering are not yet competitive with other
methods for %
$N < 300$ \cite{SpoSwa01}, so matrix multiplication is often used
by default.
Note also that the computation of a derivative or a filter using two
FTs can be combined into a single matrix multiplication. In the
example we consider here, an implementation of the FT via matrix
multiplication is found to be competitive or even faster than
FFTs for about $N\lesssim100$, see Sec.\ \ref{efficiency}. Therefore, for
the rather small $N$ that we want to consider, we focus on
the matrix multiplication method.

This leads to the second topic of the paper.
Since the proposed CFF method requires a
Yn-filter as an essential part for stability, we have to address the
implementation and efficiency of Yn-filters. We will show how
Yn-filters can be implemented by a matrix multiplication method. That
this is possible, is clear (a Yn-filter is a linear transformation of a finite
number of grid values), but we give a prescription that is
well-adapted to the present case.
Even though various software libraries for Y-transforms are in
principle available for various platforms, this is in general not true
for Yn-transforms, so a simple prescription in terms of matrices
should be of value.
As a consequence of the time-dependence of our problem, all the
required matrices for differentiation and filtering can be precomputed
at negligible startup cost, and in our case (the Einstein equations
with at least 50 variables) also at low memory cost.

As a third and final topic, we address the parallelisation of the
CFF/Yn-filter method on graphics cards (GPU computing). Concretely, we
discuss an implementation using NVIDIA's CUDA framework \cite{cuda}.
A key issue to address is that in order to avoid the bottleneck of
host-device memory transfers, it is optimal to implement the entire
calculation apart from input/output operations on a single graphics card. 
Although BLAS is available in CUDA, several
non-BLAS operations are required. 
GPU computing gives us an additional reason for a matrix
method, since on new architectures, basic linear algebra can be
expected to arrive earlier and to be better optimized than FFTs (as
was the case for CUDA during the last years).
We present some performance results for the CFF/Yn-filter method for
the single black hole test case. The non-standard feature with regard
to matrix computations on graphics cards is that the matrix
computations involve the multiplication of small-by-small matrices
with small-by-large matrices, say a $40\times40$ times a
$40\times40000$ matrix. That is, the product of small, square
differentiation and filter matrices with rectangular matrices
representing the fields with one small and one much larger dimension.
Optimization for such matrices was found to be less advanced than the
standard square matrix case using dgemm for $N\gtrsim1024$.  The
required small/rectangular matrix products achieve about $50-100$
Gflop/s, compared to $300$ Gflop/s for large matrices and a
theoretical peak around $500$ Gflop/s on the available NVIDIA
hardware. The bottom line for the GPU implementation of the black hole
example is a speed-up of a factor of about $10-20$ compared to a
single CPU implementation.

The paper is organized as follows. In Sec.~\ref{CFF}, we describe the
CFF collation method for a spherical shell, in particular the
computation of the pseudospectral derivatives.  In Sec.~\ref{filters},
we discuss the discrete transforms for the Fourier, spherical
harmonic, and spin-weighted spherical harmonic bases, and construct the
corresponding filters.  In
Sec.~\ref{numericalresults}, we discuss
various numerical features of the single black hole test case
and present some benchmarks.  We conclude in
Sec.~\ref{conclusion}. In App.~A, we summarize the
formulation of the black hole example, and App.~B gives examples for
spin-weighted spherical harmonics.

\section{Chebyshev-Fourier-Fourier collocation method}
\label{CFF}

\subsection{Coordinates and collocation grid for a spherical shell}
\label{grid}

Consider a spherical shell in three dimensions given in standard
spherical coordinates by $r\in[r_{min},r_{max}]$, $\theta\in[0,\pi]$,
and $\phi\in[0,2\pi]$. 
We introduce a discrete (Cartesian-product) grid on the shell by
\bea
	r_k &=& \frac{r_{max}+r_{min}}{2} - \frac{r_{max}-r_{min}}{2} 
                \cos \frac{\pi k}{N_r-1},
        \qquad k = 0, \ldots, N_r-1,
\label{rk}
\\
	\theta_i &=& \frac{\pi (i+\frac{1}{2})}{N_\theta},
	\qquad i = 0, \ldots, N_\theta-1,
\label{thetai}
\\
	\phi_j &=& \frac{2\pi j}{N_\phi},
	\qquad j = 0, \ldots, N_\phi-1.
\label{phij}
\eea
The radial grid is adapted to a Chebyshev spectral basis.
There are $N_r$ points in the radial direction located at the
Chebyshev extrema points plus the end points of the interval
$[r_{min},r_{max}]$. 
In latitude, there are $N_\theta$ equally spaced
points that stagger the poles at half a grid spacing. In longitude,
there are $N_\phi$ equally spaced points.

The collection of fields $u^\mu(t,r,\theta,\phi)$ on the sphere
that defines the state vector of the physical problem is represented
by the spatially discrete values
$u^\mu_{kij}(t)=u^\mu(t,r_k,\theta_i,\phi_j)$ at the collocation points.

The collocation points in the angular direction are appropriate both
for spherical harmonics, which we use for filters, and for the double
Fourier spectral basis, which we use for the computation of derivatives.
The double Fourier approach relies on periodicity in both angular
coordinates. This can be made explicit by a double covering of the
sphere, i.e.\ by doubling the range of $\theta$ by chosing
$i=0,\ldots,2N_\theta-1$ instead of $i=0,\ldots,N_\theta-1$
while keeping the grid spacing $\pi/N_\theta$ fixed. 
Equivalently, we can use the identity $(\theta,\phi) \equiv
(2\pi-\theta,\pi+\phi)$ between points on the sphere, which implies
$f(\theta,\phi)= f(2\pi-\theta,\pi+\phi)$ for any function $f$ on the
sphere. 
The fields have to be stored only for the single cover,
$\theta\in[0,\pi]$. Only when the derivatives in the
$\theta$-direction are computed, we temporarily introduce data for
$\theta\in[\pi,2\pi]$ by symmetry for convenience, so that the Fourier
derivative can be computed by the matrix multiplication
discussed below. Concretely, $N_\theta$ data points are expanded to
$2N_\theta$ points, and for the matrix multiplication we use half of
the standard matrix (a $N_\theta\times 2N_\theta$ matrix), since the
result is only needed for the single cover.

For the CFF grid, we choose an even number $N_\phi$ of points in the
$\phi$-direction, so that both $\phi_j$ and $\phi_j + \pi$ are part of
the grid. For the spherical harmonic transform required for the
filter, equal angular resolution is appropriate, so we set
\beq
  N_\phi = 2 N_\theta.
\eeq
This is also the natural choice for a physics problem that requires
roughly equal angular resolution in $\theta$ and $\phi$.  Taking into
account the staggering in $\theta$, we choose $N_\theta$ odd so that
there are points in the $x$-$y$-plane. In this case, $N_\theta=2k+1$
and $N_\phi=4k+2$ for $k$ an integer.

To illustrate that this is of course not the only way to define a
double Fourier grid, in \cite{Tic09} the $\theta$-range is
$\theta\in[0,2\pi]$, $\theta_i = \pi (2i+1)/N_\theta$ for
$i=0,\ldots,N_\theta-1$, and furthermore $N_\phi = 3 N_\theta/4$ with
$N_\theta$ a multiple of 4. The filter of \cite{Tic09} removes approximately
half the modes in the (double covered) $\theta$-direction and
one-third of the modes in the $\theta$-direction. In the present work,
similar to \cite{LinSchKid05}, such one-half or one-third rules are
not used (and apparently not crucial for stability), and hence our
grid dimensions are not adapted to such filtering.

\subsection{Cartesian tensors and smoothness}
\label{cartesian}

Since we consider not just scalar but tensor fields, we have to
discuss the smoothness of the tensor components in different coordinate
systems.
Given a tensor field with smooth components in Cartesian coordinates
$(x,y,z)$, in general its components with respect to spherical
coordinates $(r,\theta,\phi)$ are not smooth on the
$z$-axis. Spherical coordinates introduce a non-physical coordinate
singularity through the Jacobian of the coordinate transformation. One
possibility is to consider an appropriate (non-smooth) spectral basis
for spherical coordinates, for example, tensor spherical harmonics. A
simple alternative is to avoid the coordinate singularities by
computing with Cartesian tensor components on the spherical coordinate
grid (which is not uncommon in numerical relativity, e.g.\ 
\cite{LinSchKid05,Tic06,Bru99}). 
Introducing a global Cartesian coordinate system also simplifies
the treatment of varying coordinates in multiple grid domains.

For example, a vector $v^i=[v^x,v^y,v^z](x,y,z)$ in Cartesian
coordinates can be evaluated at the grid points of the spherical
coordinate grid, $x_{kij}=x(r_k,\theta_i,\phi_j)$ etc.\ As part of the
spectral method, partial derivatives are computed along coordinate
lines of spherical coordinates, that is, the spectral derivative operators
compute $\partial_r$, $\partial_\theta$, and $\partial_\phi$. However,
for the field equations the result has to be expressed in Cartesian components,
which is done using the chain rule. For the example of a vector,
\beq
\frac{\partial}{\partial x^i} v^k(\tilde x(x))
=
\frac{\partial \tilde x^j}{\partial x^i} 
\frac{\partial}{\partial \tilde x^j} v^k(\tilde x),
\label{dvdx}
\eeq
where $x^i=(x,y,z)$ and $\tilde x^i = (r,\theta,\phi)$.
The Jacobian matrix $\frac{\partial \tilde x^j}{\partial x^i}$ is known
analytically, with a pole e.g.\ in $\frac{\partial\phi}{\partial x} =
- \frac{\sin\phi}{r\sin\theta}$ at $\theta=0,\pi$ even if $r>0$ for
the shell. However, the Cartesian components $v^k(\tilde x)$ are
constant as functions of $\phi$ as $\theta\rightarrow0,\pi$, hence
$\frac{\partial}{\partial \phi}v^k(\tilde x)$ vanishes at the poles, and
the overall result is finite. In the numerical computation,
it turns out that staggering points in the $\theta$-direction so that
$\theta=0,\pi$ is not part of the grid suffices for an exponentially
convergent result. Although $\frac{\partial\phi}{\partial x}$ is
within half a grid-spacing of a pole, it is finite, and the spectral
accuracy of the numerical derivatives is sufficient for the
convergence of (\ref{dvdx}).
  
If we stored
$\tilde v^j(\tilde x) = \frac{\partial \tilde x^j}{\partial x^i} v^i(\tilde
x)$, then there would be the additional issue that the
$\frac{1}{\sin\theta}$ pole has to be differentiated numerically. To
avoid this, we could store dual vectors,
$\tilde w_j(\tilde x) = \frac{\partial x^i}{\partial\tilde x^j} w_i(\tilde
x)$, where the inverse Jacobian is finite. However, the inverse
Jacobian is multi-valued (not continuous) at the poles of the sphere,
e.g.\ $\frac{\partial x}{\partial\theta}(\theta=0)=r\cos\phi$. 
For the Y-basis this is an issue, since spectral convergence of the
expansion is lost, while the Yn-basis addresses precisely this
issue. The FF-basis does not have an immediate problem, since 
$\frac{\partial x}{\partial\theta} = r\cos\phi\cos\theta$ is 
fine as a periodic function for
$(\theta,\phi)\in[0,2\pi]\times[0,2\pi]$. We did not explore whether
the FF-basis with tensor components in spherical coordinates can lead
to spectral convergence for the tensor equations at hand, but rely
on Cartesian components and the chain rule for differentiation (\ref{dvdx}).

\subsection{Computation of derivatives in 1d}
\label{derivatives1d}

For the CFF basis, the computation of derivatives reduces to
three one-dimensional derivatives in each of the three directions.
(For CY, the spherical harmonic part is not a 1d operation.)
We compute derivatives with the matrix multiplication method, e.g.\
\cite{For98,Tre00}. For a function $f(x)$ on a 1d grid with $N$ points
$x_i$, the function values $f_i=f(x_i)$ are multiplied by a $N\times N$
differentiation matrix $D_{ij}$ to obtain the approximate derivative,
\beq
  (\partial_x f)_i = \sum_{j=0}^{N-1} D_{ij} f_j.
\eeq
For the angular directions we assume that $2N_\theta$ and $N_\phi$ are
even and that the points are equally spaced on a periodic grid, see
(\ref{thetai}) and (\ref{phij}) for the double cover. For $N$ even,
the Fourier differentiation matrix is
\beq
   FD_{ij} = \frac{(-1)^{i+j}}{2\tan(\frac{x_i-x_j}{2})} 
   \quad \mbox{for $i\neq j$},
   \quad\quad FD_{ii} = 0. %
\eeq
The Chebyshev differentiation matrix for the extrema grid
$x\in[-1,1]$, $x_i = - \cos\frac{\pi i}{N-1}$, $i=0,\ldots,N-1$, is
\beq
   CD_{ij} = \frac{c_i}{c_j}\,
   \frac{(-1)^{i+j}}{x_i-x_j} 
   \quad \mbox{for $i\neq j$},
   \quad\quad 
   CD_{ii} = - \sum_{j=0,j\neq i}^{N-1} CD_{ij},
\label{CDhat}
\eeq
where $c_k = 2$ if $k=0$ or $k=N-1$, and $c_k=1$ if $0 < k < N-1$. 
The explicit value on the diagonal is known,
but the sum in (\ref{CDhat}) is preferable for stability. 
For the radial direction, we assume the Chebyshev extrema grid
(\ref{rk}), so the differentiation matrix has to be rescaled
according to the linear transformation between $r\in
[r_{min},r_{max}]$ and $x\in [-1,1]$,
$
    \widehat{CD}_{ij} = 2 CD_{ij}/(r_{max}-r_{min}).
$
For additional details of the computation of differentiation matrices
see \cite{For98,Tre00,WeiRed00}.

We compute and store the 1d differentiation matrices of the CFF basis
once during the initialization of the time evolution.

\subsection{Computation of derivatives in 3d}
\label{derivatives3d}

For three-dimensional grids there are various options for the storage
layout of the data and for the computation of partial derivatives in
each of the three directions. We store the field values on the grid
as a one-dimensional array of size $N_{4d} = n_1 n_2 n_3 n_v$, where
$n_1=N_r$, $n_2=2N_\theta$, $n_3=N_\phi$, and $n_v$ is the number of
variables $u^\mu$, $\mu=0,\ldots,n_v-1$. The relation between
the 1d indices in (\ref{rk})--(\ref{phij}) and the linear 4d index is 
$p = k + n_1(i + n_2 (j + n_3 \mu))$.

We denote the differentiation matrices in the three spatial directions
by 
$D_1=\widehat{CD}_{n_1\times n_1}$, 
$D_2=FD_{n_2\times n_2}$,
$D_3=FD_{n_3\times n_3}$.
The basic task for differentiation given 3d data (or 4d data for
several variables) as a 1d array is to perform matrix multiplications
with a stride of 1 for the first direction, a stride of $n_1$ for the
second direction, and a stride of $n_1n_2$ for the third direction.
This is straightforward to implement, but for efficiency we want to
resort to optimized library routines. Unfortunately, BLAS for example
does not provide strided matrix-matrix multiplication. There is a
strided matrix-vector multiplication, but calling this repeatedly is
not efficient. Since our focus is on emerging computing platforms like
GPUs, choices for matrix libraries are rather limited, and hence we
look for alternative implementations.

One elegant way to proceed \cite{Tre00} is to construct 3d
differentiation matrices acting on one-dimensional arrays of size
$N_{3d}=n_1 n_2 n_3$ using the Kronecker product,
\beq
D^{3d}_1 = D_1 \otimes I_2 \otimes I_3, \quad
D^{3d}_2 = I_1 \otimes D_2 \otimes I_3, \quad
D^{3d}_3 = I_1 \otimes I_2 \otimes D_3,
\label{D3d}
\eeq
where the $I_k$ are the identity matrices of size $n_k\times n_k$, and
the $D^{3d}_k$ are of size $N_{3d}\times N_{3d}$.
The computation of the spectral derivative of a 3d field given as a 
1d vector $u$ using (\ref{D3d}) is given by the matrix multiplication
\beq
\partial_k u = D^{3d}_k u.
\label{D3du}
\eeq 
The examples in \cite{Tre00} implement the $D^{3d}_k$ as sparse
matrices in MATLAB. This leads to a very straightforward and
quite efficient implementation of (\ref{D3du}). 

The pseudospectral differentiation matrices $D^{3d}_k$ can be called
``semi-sparse''. For the remainder of this paragraph, let us set
$N_{3d}=N^3$. A dense matrix would have $N_{3d}^2=N^6$ entries.  For
finite differencing with a stencil of constant size $s$ (independent
of $N$) there are $s$ non-zero matrix elements per row for a total of
$sN_{3d}=sN^3$ elements for 3d differentiation matrices. For
pseudospectral differentiation matrices there are about $N$ non-zero
entries per row, and $N^4$ of $N^6$ elements of the 3d differentiation
matrices are non-zero. Sparse matrix libraries probably offer varying
degrees of efficiency for the semi-sparse matrices given in
(\ref{D3d}).  However, if the special sparse structure of the
$D^{3d}_k$ is not taken into account, then sparse matrix operations
are expected to be slower than strided matrix multiplication due to
the overhead in the index manipulations of the sparse matrix format
(in particular, the additional memory transfer for the index data).

The implementation that we choose uses two elementary building blocks,
BLAS matrix-matrix multiplication and a general purpose matrix
transpose. For the leading dimension of direction one, the indexing is
such that the vector $u$ containing the data for the 3d grid for
each of the variables represents a vector with $n_1n_2n_3n_v$
elements, but $u$ can also be viewed as a $n_1\times
n_2n_3n_v$ matrix. In fact,
\beq
  (u)_{n_1 n_2 n_3 n_v} =
  (u)_{n_1 \times n_2 n_3 n_v} =
  (u)_{n_1 n_2 \times n_3 n_v} =
  (u)_{n_1 n_2 n_3 \times n_v} = (u)_{n_1 \times n_2 \times n_3 \times n_v}
\label{vecmatindexing}
\eeq
as far as the memory layout is concerned, since the different matrix sizes
only refer to different ways to index the identical data.
In our implementation (C and CUDA), this
``reshape'' operation does not require any memory copies. There could
be situations where a copy operation for special memory alignment of
the rows is required, which however would be a local copy as opposed
to the non-local copies of e.g.\ a transpose operation.

The spectral derivative in the first direction can therefore be
written as the matrix multiplication
\beq
  (\partial_1 u)_{n_1\times n_2n_3n_v} =
  (D_1)_{n_1\times n_1} (u)_{n_1\times n_2n_3n_v},
\label{D1}
\eeq
where with (\ref{vecmatindexing}) the input and the result are
1d arrays of size $n_1n_2n_3n_v$.

For the derivatives in direction two and three, the data is not stored
consecutively and we cannot multiply directly by $D_2$ or $D_3$. We
implement these derivatives by performing explicit matrix transpositions.
If direction three was the last dimension, then we could consider using
some of the built-in transpose operations in BLAS and multiply by
$D_3$ from the right. However, we choose to combine all variables into
one large array $u_{n_1n_2n_3n_v}$ in order to coalesce the various matrix
operations. BLAS offers matrix multiplications with various
transposes, $AB$, $AB^T$, $A^TB$, and $A^TB^T$, 
but these are not the transposes we need. 

For the derivative in the second direction, we transpose $u$ so that
direction two becomes the leading dimension, multiply by $D_2$ from
the left, and then undo the tranpose,
\bea
  (v)_{n_2 n_3 n_v \times n_1}
  &=& (u_{n_1\times n_2 n_3 n_v})^T ,
\\
  (\partial_2 v)_{n_2 \times n_3 n_v n_1} &=&
  (D_2)_{n_2\times n_2} (v)_{n_2\times n_3n_vn_1} ,
\label{D2}
\\  
  (\partial_2 u)_{n_1 \times n_2 n_3 n_v} &=&
  ((\partial_2 v)_{n_2 n_3 n_v \times n_1})^T ,
\eea
where (\ref{vecmatindexing}) is assumed, and $u$ and $\partial_2 u$ are
1d arrays of size $n_1n_2n_3n_v$. 

Similarly, for the derivative in the third direction,
\bea
  (w)_{n_3 n_v \times n_1 n_2}
  &=& (u_{n_1 n_2 \times n_3 n_v})^T ,
\\
  (\partial_3 w)_{n_3 \times n_v n_1 n_2} &=&
  (D_3)_{n_3\times n_3} (w)_{n_3 \times n_v n_1 n_2} ,
\label{D3}
\\  
  (\partial_3 u)_{n_1 n_2 \times n_3 n_v} &=&
  ((\partial_3 w)_{n_3 n_v \times n_1 n_2})^T .
\eea
 
For the partial differential equations that we consider, we always
need all three partial derivatives. Therefore, the computation of the
derivatives as written above consists of 4 transpose operations and 3
matrix multiplications. In practice, we use CUBLAS and the
transpose from the CUDA SDK. It is likely that the transpose can be
optimized, 
but as we will see the overall performance is still dominated by the
matrix multiplication. It is interesting to note that even in the case
of a MATLAB implementation along the lines of \cite{Tre00}, using
transposes and the effectively 1d dense matrix multiplication for the
derivatives is faster than the sparse, 3d matrix implementation by
roughly a factor of 2.

In terms of the operation count, the computational kernel of the
pseudospectral CFF method as formulated above is dominated by the
matrix multiplications (\ref{D1}), (\ref{D2}), and (\ref{D3}).  They
are given by the product of a small matrix $D_k$ with a non-square
matrix $u$, $v$, or $w$ representing the data. A typical grid size for
our example is $n_1=n_2=n_3=40$ and $n_v=50$, so for the first
direction the derivative is computed as the product of a $40\times40$
matrix and a $40\times80000$ matrix. Although this is a matrix size
that is suited for parallelization, the CUBLAS 3.2 library, for example,
reaches its performance optimum for dimensions that are a multiple of
64, with double precision performance dropping from
about $300$ GFlop/s to $100$ GFlop/s if a dimension is not a proper
multiple. This is not optimal for a spectral problem where a
reasonable set of convergence runs may consist of steps
$n_1=20,24,28,\ldots,40$. The spectral method discussed here would
benefit most from the optimization of the matrix-matrix multiplication of a
small square matrix times a highly non-square matrix.

\subsection{Numerical simulations}
\label{timeevolution}

The solution of the time-dependent problem (\ref{deltu}) proceeds as
follows. First, the grid structure is initialized and all required
matrices are computed and stored. The grid does not change during the
evolution. Initial data for the physical fields $u^\mu(0)$ is
computed. 

Second, time stepping is performed by the method of lines.  We employ
a simple fourth-order Runge-Kutta (RK4) method. The allowed size of
the time-step depends on the clustering of grid points near the poles
of the spherical shell. Depending on the relative grid dimensions,
either the clustering in the $r$- or in the $\theta$-direction is more
severe. A Runge-Kutta step consists of 4 evaluations of
$S^\mu(u)-{A^{i\mu}}_\nu(u) \partial_i u^\nu$.  As part of each
substep, boundary conditions are applied.
In our example, one RK4 time step involves 600 one-directional
derivatives of individual fields plus about 10000 additional floating
point operations in the computation of the right-hand-side, where
overall the workload in the algebra is smaller than that of the
derivatives.
After one complete RK4 step, we apply the filter discussed in Sec.\
\ref{filters} to the fields.

The method is implemented in C/C++ in a package called
\textsc{BAMPS}. It inherits several features from the code
\textsc{BAM}, which is a mature infrastructure for black hole
simulations using finite differences \cite{Bru99,BruTicJan03,BruGonHan06}.

In the case of the GPU implementation, a bottleneck is the
comparatively slow memory transfer between host and device (about 30
times slower than for device-to-device copies). The
initialization step is performed on the host, and all data required
for the evolution is copied onto the device. The time evolution is
carried out completely on the device. 
In our example this is possible due to the low memory requirement of
the spectral method. If more memory is required than the device can
provide, the performance assessment changes.
Periodically, information about
the evolution is copied from the device back to the host for
processing. For simulations aimed at computing the physics of the
system, the transfer bottleneck is not a major performance limitation,
since the physical time-scale is typically much larger than the
time-step size required for numerical stability of RK4.

\section{Spherical harmonic filter for tensors}
\label{filters}

\subsection{Discrete Fourier transform as matrix multiplication}
\label{DFT}

As a first step we write the standard Fourier transform (e.g.\
\cite{Boy01}) and its inverse as matrix multiplication
transformations. Consider a real, periodic function $f(\phi)$ on the
interval $[0,2\pi]$, which is discretized by $\phi_j=\frac{2\pi}{J}j$
and $f_j=f(\phi_j)$ with $j=0,\ldots,J-1$. The backward Fourier
transform (also called Fourier synthesis or expansion in Fourier
modes) is written in terms of real Fourier modes,
\beq
f_j %
    = a_0 + \sum_{m=1}^{M-1} (a_m \cos m \phi_j + b_m \sin m \phi_j), \quad
   j = 0,\ldots,J-1.
\label{Fsynthesis0}
\eeq
The forward Fourier transform (Fourier analysis, projection onto Fourier modes)
is
\beq
	a_0 = \frac{1}{J} \sum_{j=0}^{J-1} f_j, \quad
        a_m = \frac{2}{J} \sum_{j=0}^{J-1} f_j \cos m \phi_j, \quad
        b_m = \frac{2}{J} \sum_{j=0}^{J-1} f_j \sin m \phi_j, \quad
        m = 1,\ldots,M-1. 
\label{Fanalysis0}
\eeq
When counting real degrees of freedom, the number of basis functions is
odd since $b_0$ is zero. 
This suggests using an odd number
$J=2M-1$ of sampling points in the $\phi$ coordinate so that the transform is
invertible.  However, for the
double covering of the sphere that we want to use, $J$ should be
even, so that given any $\phi_j$ the point $\phi_j+\pi$ is part of the grid. 
We therefore set
\beq
  J=2M.
\eeq
When constructing filters, invertibility is not the goal anyway.  

In matrix notation,
\bea
  f = A a + B b, && \quad 
  A_{j0} = 1, \quad A_{jm} = \cos \frac{2\pi j m}{J}, \quad
  B_{jm} = \cos \frac{2\pi j m}{J}, 
\label{Fsynthesis}
\\
  a = \tilde A f, \quad 
  b = \tilde B f, && \quad
  \tilde A_{0j} = \frac{1}{J}, \quad
  \tilde A_{mj} = \frac{2}{J} A_{jm}, \quad
  \tilde B_{mj} = \frac{2}{J} B_{jm},
\label{Fanalysis}
\eea
where $m\geq1$. The matrix dimensions are given by $f_J$, $a_M$, $b_{M-1}$,
$A_{J\times M}$, and $B_{J\times (M-1)}$.
The sine and cosine parts can be combined,
\beq
  f = C c, \quad c = \tilde C f, \quad
  c = \left( \begin{array}{c} a \\ b \end{array} \right), \quad 
  C = (A \,\,\,B), \quad
  \tilde C = \left( \begin{array}{c} \tilde A \\ \tilde B \end{array} \right),
\eeq
with dimensions indicated by $f_J$, $c_{J-1}$, and $C_{J\times(J-1)}$. 
For the numerical implementation, we precompute the transformation matrices
for the backward and for the forward transform.

\subsection{Discrete spherical harmonic transform by matrix multiplication}
\label{discreteY}

We introduce the discrete spherical harmonic transform along the lines
of \cite{SwaSpo00}, and give an implementation in terms of
matrix multiplication that relies on the pseudo-inverse of the
Legendre transformation matrix computed via the singular value
decomposition \cite{SwaSpo03}.

We consider functions on the sphere, $f(\theta,\phi)$, with the inner
product $(f,g)=\int \bar f g d\omega$, $d\omega = \sin\theta d\theta d\phi$.
The spherical harmonics are denoted by
\beq
  Y_{lm}(\theta,\phi) = \hat P^m_l(\cos\theta) e^{im\phi},
\eeq
where the $\hat P^m_l$ are normalized associated Legendre polynomials
such that $(Y_{lm},Y_{l'm'})=\delta_{ll'} \delta_{mm'}$.
We are looking for the discretized version of the backward and forward
spherical harmonic transforms,
\beq
f(\theta,\phi) = \sum_{l=0}^{\infty} \sum_{m=-l}^{l} c_{lm}
Y_{lm}(\theta,\phi),
\quad
c_{lm} = (Y_{lm},f) = \int_{S^2}\!d\omega\, \overline{Y}_{lm} f.
\eeq

We work on an equidistant two-dimensional grid of angles,
for which different choices are possible.  We choose to stagger the poles, 
and we choose an even number of points in the $\phi$ direction (because of
the Fourier double cover used for derivatives, see above). 
Setting $N_\phi=2N_\theta=2N$, there are $N\times2N=2N^2$ grid points,
\beq
  \theta_i = \frac{\pi}{N} (i + \frac{1}{2}), \quad i = 0, \ldots,N-1,
\quad\quad
  \phi_j = \frac{\pi}{N} j, \quad j = 0, \ldots, 2N-1.
\eeq

For real basis functions the discrete backward transform (synthesis,
expansion in spherical harmonics) is written as
\beq
f_{ij} = \sum_{l=0}^{L} \sum_{m=0}^{l} \hat{P}_l^m(\cos\theta_i)
(a_{lm} \cos m\phi_j + b_{lm} \sin m\phi_j).
\label{realYexp}
\eeq
We set the maximal value of $l$ (and hence also of $m$) to 
\beq
  L=N-1.
\eeq
Exchanging the order of summation according to $\sum_{l=0}^{L} \sum_{m=0}^{l} = \sum_{m=0}^{L} \sum_{l=m}^{L}$,
spherical harmonic synthesis can be written as a Legendre transform
followed by a standard Fourier transform,
\beq
f_{ij} = \sum_{m=0}^L (A_{jm} a_m(\theta_i) + B_{jm} b_m(\theta_i)), \quad 
a_m(\theta_i) = \sum_{k=0}^{L-m} (P_m)_{ik} (a_m)_k, \quad
b_m(\theta_i) = \sum_{k=0}^{L-m} (P_m)_{ik} (b_m)_k
\label{fij}
\label{backwardP}
\label{backwardF}
\eeq
where $A_{jm}$ and $B_{jm}$ are the $2N\times N$ Fourier synthesis matrices
defined in (\ref{Fsynthesis}), with $J=2N$ and $M=N$.
For each $m=0,\ldots,L$ we have defined the $N\times(N-m)$ matrix
\beq
(P_m)_{ik} = \hat{P}^m_{m+k}(\cos\theta_i)
\label{Pmatrix}
\eeq
for the Legendre synthesis, where the entries are the normalized
associated Legendre polynomials for a given $m$ evaluated for
$l=m+k=m,\ldots,L$ at the angles $\theta_i$, $i=0,\ldots,L$.

The discrete forward transform (analysis, projection onto spherical
harmonics) begins with a discrete forward Fourier transform in $\phi$,
(\ref{Fanalysis}), leading to coefficients depending on $\theta$,     
\beq
  a_m(\theta_i) = \sum_{j=0}^{2N-1} \tilde A_{mj} f_{ij}, \quad
  b_m(\theta_i) = \sum_{j=0}^{2N-1} \tilde B_{mj} f_{ij},
\label{forwardF}
\eeq
where $m=0,\ldots,L$, and $i=0,\ldots,L$. 

For each $m=0,\ldots,L$, the forward Legendre transform (analysis) is,
conceptually, the inverse of the backward transform. Written in matrix
notation, the backward Legendre transform (synthesis) from the
$(a_m)_k$ to the $a_m(\theta_i)$ in (\ref{fij}) becomes
\beq
s_N = P_{N\times(N-m)} a_{N-m}, 
\label{seqPa}
\eeq
where $s_N$ represents the synthesized $N$-vector $a_m(\theta_i)$,
$a_{N-m}$ the $N-m$-vector of coefficients $(a_m)_k$, and
$P_{N\times(N-m)}$ is the transformation matrix.

Here we encounter the usual mismatch between the number of
grid points of the rectangular $\theta$-$\phi$-grid, which is $2N^2$,
and the number of spectral coefficients $(a_m)_l$ and $(b_m)_l$, which for
$l=0,\ldots,N-1$ and $m=0,\ldots,l$ with $(b_0)_l=0$ amount only to $N^2$
coefficients.
Put differently, in general (\ref{seqPa}) cannot be inverted since for
$m>0$ the matrix $P_{N\times(N-m)}$ is not even square. There are $N$
equations for $N-m$ unknowns $a_{N-m}$. 

However, we can compute the analysis $a=(P,s)$ by a sum over grid
points, which looses information, so that $\tilde s = P a$ is an
approximation of $s$. 
For the Gaussian collocation points of the Legendre functions (which
we do not use), all that would be needed are appropriate weights $w_i$ for
$a_k=\sum_i w_i P_{ik} s_i$. For a general set of collocation points,
we can define an (in general non-diagonal) weight matrix $W$ so that
$a = P^T W s$, see for example \cite{SwaSpo00}, which also discusses
clever ways to compute and store $W$ and/or $P^T W$. 

In principle, one could generalize \cite{SwaSpo00} or the method based
on special collocation points to spin-weighted spherical harmonics, at
the cost of increased analytic complexity. However, especially in the
context of a matrix method, there is a straightforward alternative.
As a simple, direct way to invert $s=Pa$ in the appropriate manner, we
follow \cite{SwaSpo03} and note that we can define
\beq 
\tilde a_{N-m} = P^+_{(N-m)\times N} s_N,
\label{forwardL}
\eeq
where $P^+$ denotes the Moore-Penrose pseudo-inverse of the matrix
$P$.

The pseudo-inverse $A^+$ of a real matrix $A$ is the unique matrix
satisfying $AA^+A=A$, $A^+AA^+=A^+$, $(AA^+)^T=AA^+$, and
$(A^+A)^T=A^+A$, cmp.\ \cite{BenGre03}.  For example, the first
relation means that although $AA^+$ is in general not the identity, it
still maps $A$ to $A$.  $AA^+$ is the orthogonal projector onto the
space spanned by the columns of $A$.  If the inverse exists, then
$A^+=A^{-1}$.
The fact we need here is that even if we cannot solve a linear
equation $Ax=b$ because the inverse of $A$ does not exist, we can
still look for vectors $x$ that minimize $\|Ax-b\|$. There may be
several such vectors. The pseudo-inverse defines the unique vector $x
= A^+b$ that minimizes $\|A x-b\|$ and has the smallest norm $\|x\|$.

The pseudo-inverse can be computed using the singular value
decomposition (SVD) of $A$, $A=USV^T$. Here $U$, $S$, $V$ are
matrices, and in particular $S$ is diagonal (and in general
non-square). For this decomposition, we have $A^+=VS^+U^T$, and the
pseudo-inverse of $S$ is obtained by taking its transpose and
replacing non-zero entries $S_{ii}$ by $1/S_{ii}$.

In summary, the pseudo-inverse allows us to define the forward
Legendre transform (\ref{forwardL}) as the least-squares approximation
to the inverse of the backward transform via the pseudo-inverse. 
Written out in components, 
the forward Legendre transform of the discrete spherical harmonics
transform is
\beq
(a_m)_k = \sum_{i=0}^{L} (P_m^+)_{ki} a_m(\theta_i), \quad
(b_m)_k = \sum_{i=0}^{L} (P_m^+)_{ki} b_m(\theta_i),
\eeq
with $a_m(\theta_i)$ and $b_m(\theta_i)$ obtained from the forward
Fourier transform, (\ref{forwardF}).

We can choose to precompute and store the matrices $P_{N\times(N-m)}$ and
$P^+_{(N-m)\times N}$ for each $m$.  Since this is done once at
startup time, parallelization of the SVD routine is not an issue.
We use the GSL for Legendre polynomials and the SVD \cite{GalDavThe09}.

\subsection{Discrete spin-weighted spherical harmonic transform by matrix multiplication}
\label{Yn}

Spin-weighted spherical harmonics are a generalization of spherical
harmonics. The spin weight refers to how a given function on the
sphere transforms under the rotation of basis vectors.  Spin-weighted
spherical harmonics were first discussed in terms of spin raising and
lowering operators in \cite{NewPen66,GolMacNew67}, which also leads to
a definition in terms of Wigner $d$-functions.
Any tensor of degree $k$ on the sphere can be naturally decomposed as
a linear combination of tensor spherical harmonics, which are products
of the basis vectors with the spin-weighted spherical harmonics
\cite{GolMacNew67,Cam71}, see Sec.~\ref{spindecomp}.

Here we use the definition given in \cite{WiaJacVie05}, see also
\cite{KosMasRoc00,WiaJacVan05}. A spin-$n$ function on the sphere,
$f(\theta,\phi)$, transforms under a basis rotation by an angle
$\psi$ according to $f = e^{-in\psi} f$. The sign convention
for $n$ is opposite to the spin weight $s=-n$ defined in
\cite{NewPen66,GolMacNew67,Cam71}, which however does not matter for
filters constructed as a forward-backward transform.
The spin-weighted spherical harmonics, $Y^n_{lm}(\theta,\phi)$, are
spin-$n$ functions on the sphere for a given $n$. They form an
orthonormal basis in the space of spin-$n$ functions with
orthonormality and completeness relations
\bea
\int_{S^2} d\omega \overline{Y^n_{lm}}(\omega) Y^n_{l'm'}(\omega)
&=&
\delta_{ll'} \delta_{mm'},
\\
\sum_l \sum_{|m|\leq l} 
\overline{Y^n_{lm}}(\omega') Y^n_{lm}(\omega)
&=&
\delta(\omega',\omega),
\eea
where $\omega=(\theta,\phi)$ and 
$\delta(\omega',\omega)=\delta(\cos\theta'-\cos\theta) \delta(\phi'-\phi)$.
Hence, any spin-$n$ function on the sphere is uniquely given by
\beq
f(\omega) = \sum_l \sum_{|m|\leq l} c^n_{lm} Y^n_{lm}(\omega),
\quad
c^n_{lm} = (Y^n_{lm},f) = 
\int_{S^2} d\Omega \overline{Y^n_{lm}}(\omega) f(\omega).
\label{Gncn}
\eeq
In the above $l$ is assumed to be equal to or larger than $|n|$, which is
implemented with the convention that 
\beq
  Y^n_{lm}(\omega) = 0 \quad\mbox{and}\quad c^n_{lm}=0 
  \quad \mbox{if $l<|n|$ or $l<|m|$}.
\eeq
The definition of the spin-weighted spherical harmonics (see below) gives
\beq
  \overline{Y^n_{lm}}(\omega) = (-1)^{n+m} Y^{-n}_{l(-m)}(\omega).
\label{Ynnegative}
\eeq
Spin-$0$ corresponds to the standard, non-weighted spherical harmonics,
$
  Y^0_{lm}(\omega) = Y_{lm}(\omega),
$
for which we have the standard orthonormality and completeness relations as a
special case of the relations above.

For the numerical computation of the spin-weighted spherical harmonics
we use recursion formulas, as opposed to the non-recursive definition
of the Wigner $d$-functions or the spin operators that are also given
in \cite{WiaJacVie05}. There are different ways to express the
$Y^n_{lm}$ in terms of the $Y_{lm}$, depending on which recursion
relation is used for the $\theta$-derivative of the associated
Legendre polynomials, compare \cite{WiaJacVie05,KosMasRoc00}.  While
\cite{KosMasRoc00} is simpler in the $\theta$-dependence of the
coefficients, \cite{WiaJacVie05} is simpler in the range of $l$, in particular
for band-limited functions on a given grid.
(We note in passing that the coefficients for negative spin weight in
\cite{KosMasRoc00} have to be corrected since the normalization of the
spin-weighted spherical harmonics is non-standard and
(\ref{Ynnegative}) does not hold.) 
A few simple examples can be found in App.~B.

The basic recursion formula employed in \cite{WiaJacVie05} is
\beq
Y^n_{lm} = 
\alpha_{nl} 
\frac{\frac{m}{l} - \cos\theta}{\sin\theta} 
Y^{n-1}_{lm}
+ \beta_{nlm} \frac{1}{\sin\theta} Y^{n-1}_{l-1,m}
\label{Yninc}
\eeq
for decreasing $n$, and for increasing $n$ it is
\beq
Y^n_{lm} = 
\alpha_{(-n)l} \frac{\frac{m}{l} + \cos\theta}{\sin\theta} Y^{n+1}_{lm}
- \beta_{(-n)lm} \frac{1}{\sin\theta} Y^{n+1}_{l-1,m} ,
\label{Yndec}
\eeq
with coefficients
\beq
  \alpha_{nl} = \left( \frac{l-n+1}{l+n} \right)^{\frac{1}{2}}, 
\quad
  \beta_{nlm} = \frac{1}{l} \left(  
     \frac{2l+1}{2l-1} \frac{(l+n-1)(l^2-m^2)}{l+n}
  \right)^{\frac{1}{2}} .
\label{Yncoeffs}
\eeq
As before, $Y^n_{lm} = 0$ for $l<\mbox{max}(|m|,|n|)$. 

This is a two-term recursion in $l$. Since the coefficients are
functions of $\theta$, the integration for analysis changes. 
Starting with $n>0$, there are $n+1$ terms involving 
$Y^0_{lm}$ multiplied by $\cot^p\theta/\sin^q\theta$ with $p+q=n$.
However, the overall behavior at the poles is regular.
Furthermore, since we stagger the
grid no extra measures at the poles should be necessary. 
The result of the recursion can be written
\beq
  Y^n_{lm}(\theta,\phi) = 
  \sum_{p=0}^n \gamma^n_{plm}(\theta) Y_{(l-p)m}(\theta,\phi).
\label{YnlmFromYlm}
\eeq
The usual way to proceed is to compute the expansion coefficients with
respect to the $Y^n_{lm}$ using some existing
implementation of the spherical harmonic transform.
The coefficients in (\ref{YnlmFromYlm}) depend on $\theta$,
which means when considered as functions of $\theta$ the terms of the
expansion are {\em not} spherical harmonics. However, when computing the
transform we can move the additional $\theta$ dependence into the
function that is to be transformed, e.g.\
$
   (\frac{Y_{lm}}{\sin\theta}, f) =  (Y_{lm}, \frac{f}{\sin\theta}).
$
As a result, the $Y^n_{lm}$-transform is computed as the linear
combination of $|n|+1$ $Y_{lm}$-transforms of the rescaled function
$f$. 

In our application we implement the spin-$n$ spherical harmonic
transform as a matrix multiplication (in particular since $l$ is
appropriately small). Rather than computing $|n|+1$ spin-0 transforms
based on (\ref{YnlmFromYlm}), we use the recursion
(\ref{Yninc})--(\ref{Yncoeffs}) directly to compute a single
transformation matrix for the Legendre-part of the transform. For $|n|
\leq 3$, this avoids a factor of up to 4 in the number of transforms.

Analytically, when computing (\ref{Yninc})--(\ref{Yncoeffs}) or
(\ref{YnlmFromYlm}) it does not matter which type of recursion is used
(\cite{WiaJacVie05} or \cite{KosMasRoc00}).  However, when computing
associated Legendre polynomials from standard Legendre polynomials
numerically, certain recursions in $l$ are stable, while some
recursions in $m$ are not
as stable. To our knowledge a corresponding large $n$ study has not
been carried out for spin-$n$ spherical harmonics and different
recursions. But note that in our case $n$ corresponds to the
tensor-degree of the physical fields and is therefore a small, fixed
number (that in particular does not increase like $m$ and $l$ when
increasing the accuracy of the spectral approximation).
Still, the numerically implementations may differ in accuracy.

More importantly, we have to ask whether the pseudo-inverse
method is applicable to the computation of the analysis matrices. 
That the pseudo-inverse exists is more or less clear, since
for each $n$ we have the same orthogonality and completeness relations that
hold for the $n=0$ case. 
Numerically, it is not clear a priori how well the
pseudo-inverse/SVD algorithm for the analysis matrices handles the
differences in the $\theta$-dependence.

We summarize the actual computation. For spin-weighted spherical harmonics
we define
\beq
  Y^n_{lm}(\theta,\phi) = \hat P^n_{ml}(\theta) e^{im\phi},
\eeq
where the $\hat P^n_{ml}$ are directly related to the Wigner $d$-functions,
$\hat P^n_{ml}= (-1)^n \sqrt{\frac{2l+1}{4\pi}} d^l_{m(-n)}$. These
``spin-$n$ associated Legendre polynomials'' are computed by the
recursion formulas (\ref{Yninc})--(\ref{Yncoeffs}).  In principle we
are looking for a numerical implementation of the Wigner
$d$-functions, but this is not readily available on most platforms.
Given a code-library function for the computation of the normalized
associated Legendre polynomials $\hat P^m_l(\theta_i)$, the
recursion formulas are directly implemented by recursive function
calls that increase or decrease $n$ until $n=0$. 
In our case, $n = -3,\ldots,+3$, with $n$ an integer.
For $n<0$, we can also use
\beq
  \nP^{n}_{lm} = (-1)^{n+m} \nP^{-n}_{l,-m}.
\eeq
The result is the $N\times(N-m)$ synthesis matrix
\beq
  (P^n_m)_{ik} = \hat P^n_{(k+m)m} (\theta_i)
\eeq
for each $m$ and $n$, in analogy to the spin-0 case, (\ref{Pmatrix}).
For the spectral analysis we use the pseudo-inverse 
\beq
(Q^n_m)_{ki} = ([P^n_m]^+)_{ki}
\eeq
of $(P^n_m)_{ik}$, where as for the spherical harmonics
$k=0,\ldots,L-m$ and the $(Q^n_m)_{ki}$ are $(N-m)\times N$ matrices.

The spin-weighted spherical harmonic transform defines a projection
filter $F_n(f)$ for functions $f$ of definite spin-weight $n$. Given
$f$, we compute the discrete forward transform followed by the
discrete backward transform for some finite $l\leq L$,
cmp.~(\ref{Gncn}). Note that $F_n$ is a linear operation, and using
(\ref{Ynnegative}) we have 
\beq 
\overline{F_n(f)} = F_{-n}(\bar f).
\label{ccFn}
\eeq
For $n\neq0$ we have for non-trivial $f$ that $F_n(f)\neq F_{-n}(f)$,
so even if $f=\bar f$ we have $\overline{F_n(f)}\neq F_{n}(f)$.
Hence, even if $f$ is real, in general the projection $F_n(f)$ is complex.

The discrete spin-$n$ spherical harmonic transform and the
corresponing filter is computed in complete analogy to the standard
($n=0$) case.  The matrices $P^n_m$ and $Q^n_m$ are precomputed. For
our main application we only store the filter matrix $F^n_m(n_f)$ as
defined in (\ref{FasPfQ}) of Sec.~\ref{filterprojection} on filters. 
We need $|n|\leq3$ and $0\leq m\leq N$. During the evolution of the
physical fields, the filter is computed by computing the discrete
Fourier analysis (\ref{forwardF}), followed by the discrete spin-$n$
associated Legendre projection (\ref{FasPfQ}), followed by the
discrete Fourier synthesis (\ref{backwardF}). The Fourier transforms
are independent of $n$.

\subsection{Spin-weight decomposition of tensors 
  with respect to a tetrad or triad}
\label{spindecomp}

In preparation for the construction of spin-weighted filters,
we decompose tensors according to their spin weight.
Consider Minkowski space with coordinates $(t,x,y,z)$ and metric
$\eta_{ab}=\mbox{diag}(-1,1,1,1)$. 
We also consider basis vectors aligned with spherical coordinates
$(t,r,\theta,\phi)$, but with components in the Cartesian basis
$(t,x,y,z)$.  We define the right-handed, orthonormal tetrad
$(t^a,r^a,\theta^a,\phi^a)$ by
\beq
\begin{array}{ll}
t_a = (-1, 0, 0, 0),
&
\theta_a = (0, \cos\theta\cos\phi, \cos\theta\sin\phi, -\sin\theta),
\\
r_a = (0, \sin\theta\cos\phi,\sin\theta\sin\phi,\cos\theta), 
&
\phi_a = (0, -\sin\phi, \cos\phi, 0).
\end{array}
\label{tetrad}
\eeq
The basis vectors
tangential to the coordinate spheres are replaced by the two complex vectors
\beq
m_a = \frac{1}{\sqrt{2}}(\theta_a + i\phi_a), \quad
\barm_a = \frac{1}{\sqrt{2}}(\theta_a - i\phi_a),	
\eeq
where $\barm_a$ is the complex conjugate of $m_a$. 
The orthonormality relation of the complex tetrad
$e^a_\mu=(t^a,r^a,m^a,\barm^a)$ with respect to the Minkowski metric
is
\beq 
t^at_a=-1, \quad r^a r_a = 1, \quad m^a\barm_a = 1, \quad \barm^a m_a = 1,
 \quad\text{others zero.}
\eeq
In terms of $e^a_\mu$, orthonormality and completeness read
$\eta_{ab} \overline{e}^a_\mu e^b_\nu = \eta_{\mu\nu}$ and
$\eta^{\mu\nu} \overline{e}^a_\mu e^b_\nu = \eta^{ab}$.
Introducing the conjugate dual of the complex tetrad, 
$f^\mu_a = \eta^{\mu\nu}\eta_{ab} \overline{e}^b_\nu$, this becomes
$f^\mu_a e^a_\nu = \delta^\mu_\nu$ and
$f^\mu_a e^b_\mu = \delta^a_b$.

Any tensor on Minkowski space can be written in terms of the complex
tetrad.
For a vector $v^a$,
the expansion is $v^a = \tilde v^\mu e^a_\mu$ with coefficients
$\tilde v^\mu = (e_\mu,v) = \eta^{\mu\nu}\eta_{ab}\overline{e}^a_\nu
v^b = f^\mu_a v^a$. For $e^a_\mu=(t^a,r^a,\theta^a,\phi^a)$,
this can be written as
\bea
v^t &=& - t_a v^a, \quad  v^r = r_a v^a, 
\quad v^m = \barm_a v^a,
\quad v^\barm = m_a v^a, 
\\
v^a &=& v^t t^a + v^r r^a + v^m m^a + v^\barm \barm^a.
\label{vtetraddecomp}
\eea
A tensor of degree $k$ is expanded as
\beq
  \tilde T^{\mu_1\ldots\mu_k} = 
  f^{\mu_1}_{a_1} \ldots f^{\mu_k}_{a_k} T^{a_1\ldots a_k},
\quad
  T^{a_1\ldots a_k} = \tilde T^{\mu_1\ldots \mu_k} 
  e^{a_1}_{\mu_1} \ldots e^{a_k}_{\mu_k}.
\label{tensorexpansion}
\eeq
This construction simplifies trivially to the case of
three-dimensional Euclidean space by dropping $t^a$ and replacing the
indices by $i=1,2,3$ and $\mu=1,2,3$.

The key property of the complex tetrad that concerns us here is its
transformation under rotations about a given radial direction $r^a$. 
The vectors $r^a$ and $t^a$ do
not change. The vector $m_a$ is chosen
for its simple transformation under such rotations,
\beq
	m'_a = e^{i\psi} m_a,
\label{mprime}
\eeq 
where $\psi$ is the angle of the rotation.
The spin weight of a function $f$ constructed from a tensor by
contractions with the tetrad refers to its behavior under rotations
of the tetrad vectors. %
If such a function transforms under tetrad rotations as
\beq
       f' = e^{-i\psi \mys} f,
\label{Tprime}
\eeq
we call it a function with spin-weight $\mys$. 
Referring to
(\ref{vtetraddecomp}), we have $\mys(v^r)=0$, $\mys(v^m)=+1$, and
$\mys(v^\barm)=-1$.
For the tetrad vectors themselves, we define
$\mys(t^a) = \mys(r^a) = 0$, $\mys(m^a) = -1$, and $\mys(\barm^a) = +1$.
According to (\ref{Tprime}), for products of
spin-weighted functions we
have $\mys(f_1f_2)=\mys(f_1)+\mys(f_2)$. For products of tetrad
vectors, 
$\mys(e^{a_1}_{\mu_1} \ldots e^{a_k}_{\mu_k} ) 
  = \sum^k_{j=1} \mys(e^{a_j}_{\mu_j})$. For example, $r_ar_b$,
$r_am_b$, $m_am_b$, and $m_a\barm_b$ have spin-weights $0$, $-1$, $-2$,
and $0$, respectively.
Products of tetrad vectors have a well defined spin-weight, but the
sum of spin-weighted tensors is in general a tensor without
well-defined spin weight. 

Note the distinction between coordinate rotations and tetrad
rotations. By definition, any tensor is covariant under coordinate
transformations, but here we have introduced additional structure, the
tetrad, and discuss how functions that are constructed from the tetrad
and tensors transform when the tetrad is transformed.  
The physics of the problem we consider is rotation invariant, i.e.\ it
does not refer to a preferred choice of $z$-axis or tetrad vector
$m^a$. Concretely, if $m^a$ is not part of the construction of a
physical field $v^a$, then $\mys(v^a) = 0$. If we choose to expand $v^a$
in terms of the tetrad, then its components acquire specific spin
weights, but each term of the sum in $v^a=v^\mu e^a_\mu$ has spin-weight 0.

\subsection{Filters defined by spherical harmonic projection}
\label{filterprojection}

In this work, the main application of the discrete (scalar and
spin-weighted) spherical harmonics transform is its use as a filter.
It is unclear a priori what type of filtering is needed or optimal for
the Einstein equations implemented with the particular CFF method that
we consider, and any filtering scheme has to be carefully evaluated.

First of all, in order to suppress high-frequency modes near the
poles, we expand $f_{ij}$ by the forward transform in spherical
harmonics up to degree $L$, i.e.\ we project onto the spherical
harmonics basis. The backward transform results in an approximation
$\tilde f_{ij}$ of the original $f_{ij}$ with equiangular resolution
over the sphere, which in particular means that the high frequencies
that can be represented on the $\theta$-$\phi$ grid but are unwanted
near the poles have been eliminated. This removes certain restrictions
on the time step size due to clustering of points near the poles in
the $\phi$-direction. Transforming to and from spherical harmonics for
a finite $L$ defines a projection filter. In the context of the
double Fourier spectral method on the sphere (for scalar fields), the
projection filter ensures equivalence to the more standard spherical
harmonics method to compute derivatives.

We assume that $L$ is the maximal degree of spherical harmonics
represented on the grid, and we define additional filtering by
explicitly removing the top $n_f$ of the highest degree $l$-modes,
i.e.\ $l\leq L-n_f$ . In our case there are two unrelated reasons to
do so. For non-linear problems, there is a large variety of approaches
\cite{Boy01} to deal with the non-linear mode mixing.  For example,
for quadratic non-linearities the two-thirds rule can be helpful for
one-dimensional intervals, while on the sphere this may become a
one-half rule since the basis is not 'reflective'.
It is unclear a priori what type of filtering is needed or optimal for
the Einstein equations, which are worse than quadratically non-linear.
As in \cite{KidLinSch04,LinSchKid05}, but in contrast to \cite{Tic09}
which uses a different formulation of the Einstein equations, we do
not resort to filtering one-half or one-third of the modes. This does not 
appear
to be necessary, neither in the radial nor in the angular direction,
and we have not investigated this here. However, the residual
linear growth discussed in Sec.~\ref{numblackhole} might be addressed with
additional filtering (or alternatively by improved boundary conditions). 

A second issue is the tensor character of the fields in combination
with the Cartesian coordinates. This leads to the observation that a
small $n_f > 0$ is required for stability (here $n_f=4$), which
depends on the rank of the tensors but not on the grid size (e.g.\
$n_f=N/3$ for filtering the top third).
To examine the Cartesian tensor issue, we consider three types of
filters based on scalar and spin-weighted spherical harmonics, which
we call the scalar Y-filter, the tensor Yn-filter, and the graded
Yg-filter.
For the Y-filter, we apply the standard, non-weighted spherical
harmonics filter to each field $u^\mu$, ignoring the tensor character
of the fields. The Y-filter addresses some of the clustering issues of
the double Fourier method, but there remains a strong instability,
which however appears to be cured when the Yn-filter using projection
onto spin-weighted spherical harmonics is used, see
Sec.~\ref{numblackhole}.

One view of the problem is that Cartesian components
introduce additional angular dependence compared to spherical
coordinates, which effectively increases the order of a spherical
harmonic expansion by one for each spatial tensor index.
Consider the spherically symmetric scalar function $f(x,y,z)=r$, which
requires only $l=0$ in a spherical harmonic expansion.  Its first
derivative $\partial_xf=x/r=\sin\theta\cos\phi$ is the component of a
Cartesian vector, which corresponds to $l=1$.  Its
second derivative
$\partial_x\partial_xf=\frac{1}{r}(1-\frac{x^2}{r^2})$, which is the
component of a 2-tensor, requires $l=0$ and $l=2$.
Analogously, referring to (\ref{tetrad})--(\ref{tensorexpansion}),
each contraction with $m_i$ to compute components of a tensor in the
spherical basis multiplies the Cartesian component by a first order
polynomial in $\sin\theta$ etc., which increases the $l$ required in a
spherical harmonic basis by one. Since on the numerical grid we can
only represent a finite, maximal degree $L$, for each
spatial index of a tensor in Cartesian components the available degree $L$ is
effectively lowered by one compared to spherical coordinates. 

Let us denote by $d(u^\mu)$ the spatial degree of the Cartesian tensor
component, i.e.\ the number of spatial indices of the
variable $u^\mu$. For example, $d(g_{tt})=0$, $d(g_{tx})=1$, $d(g_{xx})=2$.
Then the effective maximal degree $L_{\eff}$ represented on the grid is 
$L-d(u^\mu)$.  
In the evolution equations and the constraints, the different spatial
degrees are coupled, e.g.\ $\myPhi_{ijk} \simeq \partial_ig_{jk}$. 
This suggests that the tensor Yn-filter should be used with $n_f\geq3$
so that each spin-weight mode is representable at the same maximal order
$L_{\eff}$ on the grid. 

On the other hand, from this point of view the scalar Y-filter is
problematic, since for a given $n_f$ it does not project onto
a basis at the same $L_{\eff}$. For example, suppose we want to
project some given 3d data onto spherically symmetric data. If we
choose the Y-filter with $n_f=L$, then a scalar function is
correctly projected onto its spherically symmetric monopole,
but for a vector we have to use $n_f=L-1$, and e.g.\ for $\myPhi_{ijk}$ 
it should be $n_f=L-3$. This leads us to consider an improved version
of the scalar filter, which was also (and possibly for the first time)
considered in \cite{Tic09}. 
We define
\beq
  n_f(\mu) = n_f - d(u^\mu),
\label{nfofmu}
\eeq
and introduce what we call a ``graded'' Y-filter, or Yg-filter,
where the top $n_f(\mu)$ components in the Y-basis are zeroed.
The Yg-filter improves on the Y-filter since the highest order
$l$-modes are now treated consistently across the different spatial
ranks of the tensor components. However, since the Cartesian
components are actually a mixture of different spin-weights, compared
to the Yn-filter the Yg-filter does not treat the intermediate
spin-weights correctly. As we show in Sec.~\ref{numblackhole}, the
Yg-filter cures one type of instability present in Y-filter
simulations.  Yet an additional, more slowly growing instability is
left over, which however the Yn-filter is able to handle.

The actual implementation of the filters is as follows.
Given a general tensor, we cannot apply the Yn-filter directly. First,
the tensor is decomposed according to (\ref{tensorexpansion}). Each
component function in the expansion is filtered according to its spin
weight.  The result is recombined again as in (\ref{tensorexpansion}).
Note that (\ref{ccFn}) is compatible with the $m^i$ and
$\barm^i$ vectors of the tetrad.
For the tetrad components of a real vector $v^i$, 
$
  \overline{F_n(v^m)} = F_{-n}(v^{\barm}),
$
which is just as it should be since the spin-weights of $v^m$ 
and $v^{\barm}$ have opposite sign.
Denoting the general filter operation by $F$
and the specific spin-$n$ version by $F_n$, we
have for example $F(g_{tt}) = F_0(g_{tt})$ and $F(g_{tm})=F_{1}(g_{tm})$.
With (\ref{ccFn}) and linearity of $F$, we can reexpress the filter 
operation in terms of non-complex basis vectors. For example,
\beq
  F(g_{t\theta}) 
  = \frac{1}{\sqrt{2}} (F_1(g_{tm}) + F_{-1}(g_{t\barm}))
  = \sqrt{2} Re(F_1(g_{tm}))
  = Re(F_1(g_{t\theta}) + i F_1(g_{t\phi})).
\eeq

The projection filter is implemented as a forward Fourier
transform in the $\phi$-direction, followed by the projection filter
\beq
\tilde s_N = (PP^+)_{N\times N} \, s_N
\eeq
onto the Legendre basis for each $m$ in the $\theta$-direction,
followed by a backward Fourier transform in the $\phi$-direction. 
Here $P$ and $P^+\equiv Q$ refer to the matrices appropriate for either
the standard or spin-weighted Legendre transforms.
This can be generalized to additional filtering by
\beq
\tilde s_N = F_{N\times N} s_N, \quad
F_{N\times N} = P_{N\times(N-m)} f_{\text{diag}(N-m)} P^+_{(N-m)\times
N},
\label{FasPfQ}
\eeq
where the elements of the diagonal matrix $f$ are one for modes that
are to be maintained and zero for the $n_f$ or $n_f(\mu)$ modes that
are to be removed. The standard
choice considered in the literature is to remove the top 4 modes
\cite{KidLinSch04,LinSchKid05}, in which case
$f=\text{diag}(1,\ldots,1,0,0,0,0)$. Recall that $P_{N\times(N-m)}$
stands for $(P_m)_{ik}$ with $k=0,\ldots,L-m$, with the entries
obtained for $l=m+k=m,\ldots,L$. Therefore, zeroing the top 4
components of the $(N-m)$-vector $P^+_{(N-m)\times N} s_N$
removes the components $l=L-3,L-2,L-1,L$. 

If storage is not an issue, we can precompute and store the different
$F_{N\times N}$ for each $m$, which requires $O(N^3)$ storage.  For
the Einstein equations in GHG form on a spherical shell, storage is
not much of an issue since the filter is applied to each of about 50
variables for every value of the radius, and $F_{N\times N}$ is
independent of $r$. In our example, storing $F_{N\times N}$ is roughly
equivalent to requiring storage for 51 instead of 50 variables, with
somewhat more storage required if the matrices for each spin-weight
are stored.
\section{Numerical results}
\label{numericalresults}

In this section we first 
present numerical experiments for the single black hole test case in
Sec.~\ref{numblackhole}, 
and then evaluate the computational efficiency of the
pseudospectral matrix method in Sec.~\ref{efficiency},

\subsection{Test case of a single, evolving black hole}
\label{numblackhole}

As a non-trivial application of the CFF/Yn-filter method, we consider
the basic example of a static, spherically symmetric single black
hole.  Analytically, the time derivatives $\partial_t u^\mu(t,x,y,z)$
all vanish. The discretization error of the numerical method leads to
a non-trivial time evolution, which in particular can depart from
spherical symmetry. The numerical method is successful if the system
settles down in a stable stationary state of the discretized equations
that approximates the analytical solution, where all the $\partial_t
u^\mu$ have dropped to the level of the round-off error.

We discuss a set of time evolutions on a single spherical shell. 
The initial data is the same in each case, see %
\ref{initialdata},
but approximated on different grids of size $N_r\times N_\theta\times N_\phi$.
For all runs discussed here, the radial coordinate extends from $r=1.8$
to $11.8$, which we label configuration R10.
We begin our discussion with
examples that are numerically stable for long times, i.e.\ the full
CFF/Yn-filter method, and then discuss the effect of using different
filters and different time step sizes.
Part of the default configuration is the Yn-filter with $n_f=4$ and
a time step with $\lambda = \Delta t/\Delta x_{\min} = 4.0$ (see
below).

\begin{figure}[t]
  \centering
  \includegraphics[width=80mm]{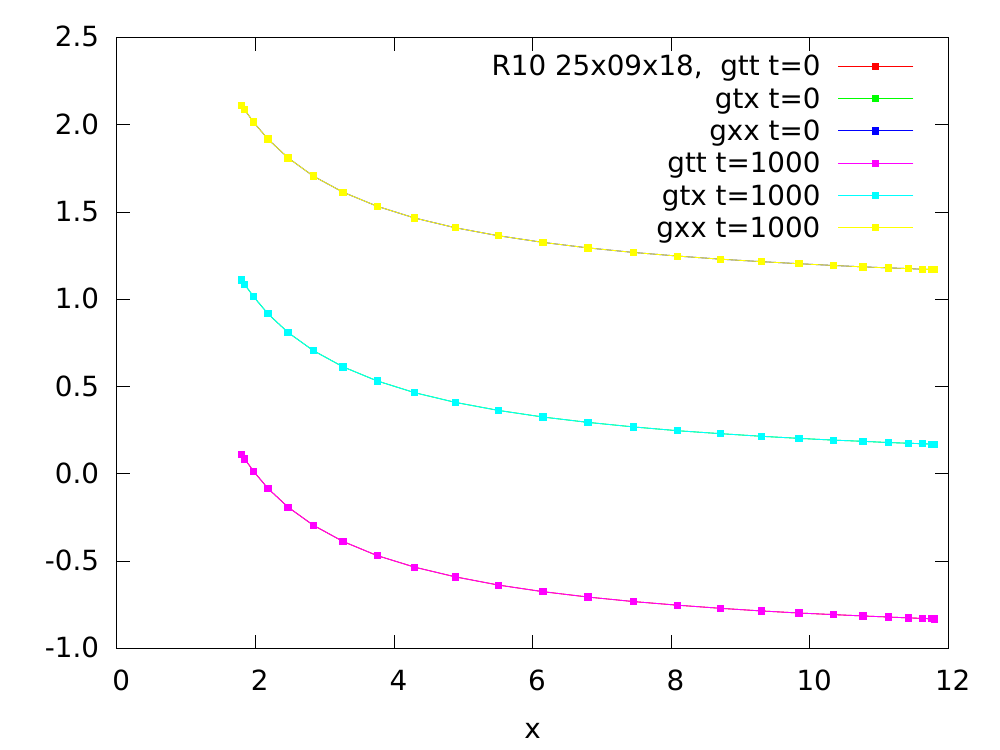} 
  \includegraphics[width=80mm]{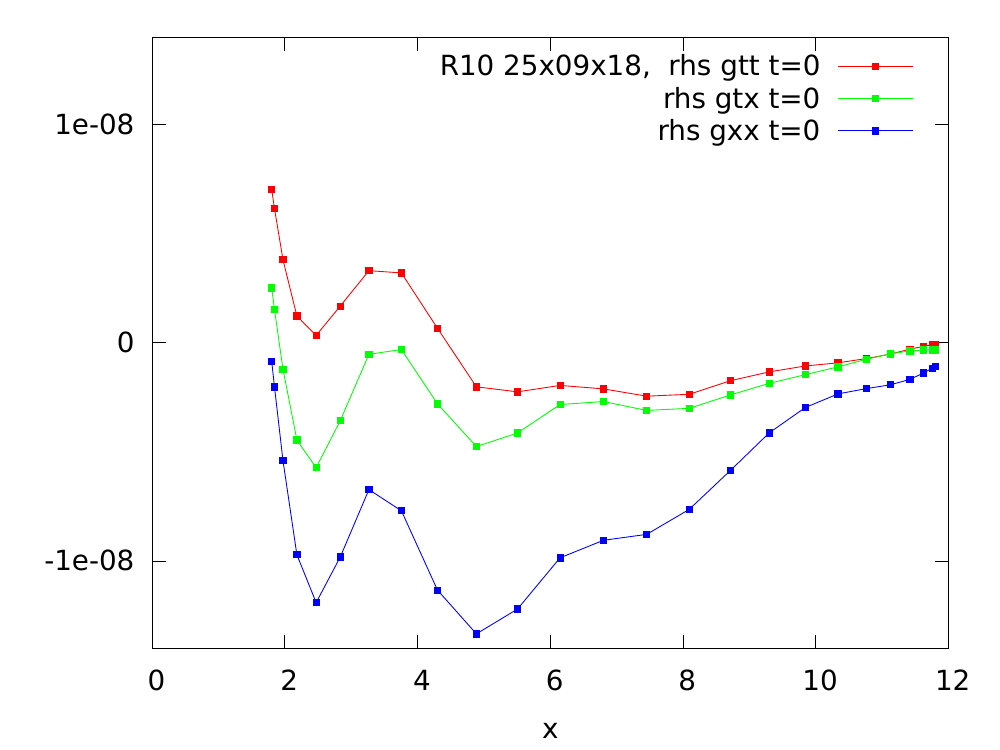} \\
  \includegraphics[width=80mm]{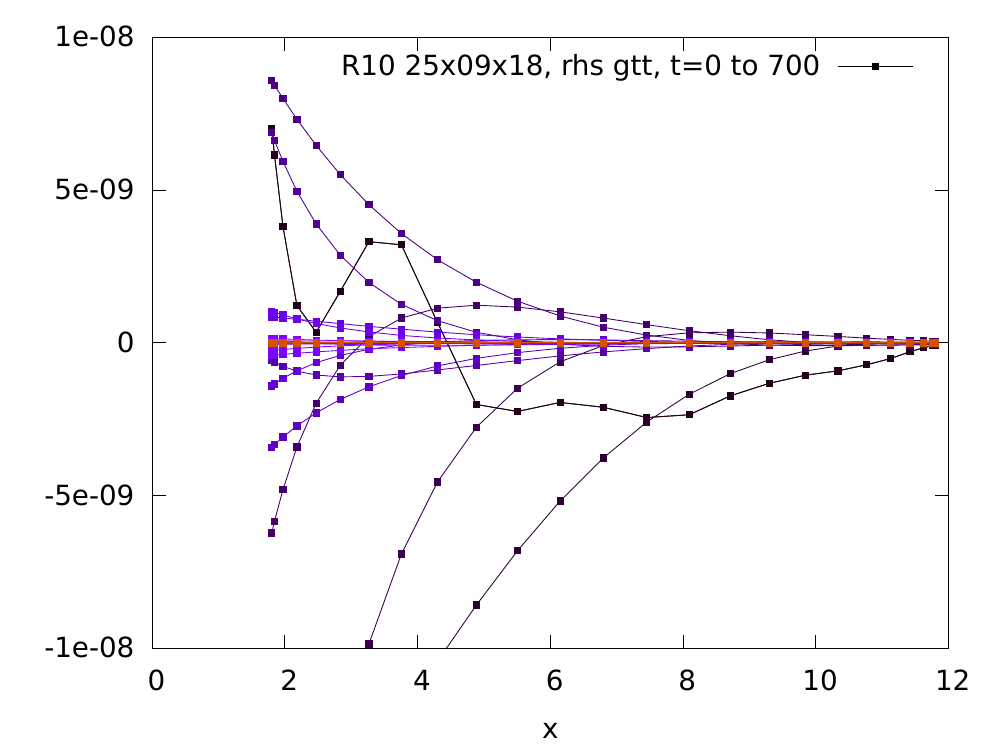}
  \includegraphics[width=80mm]{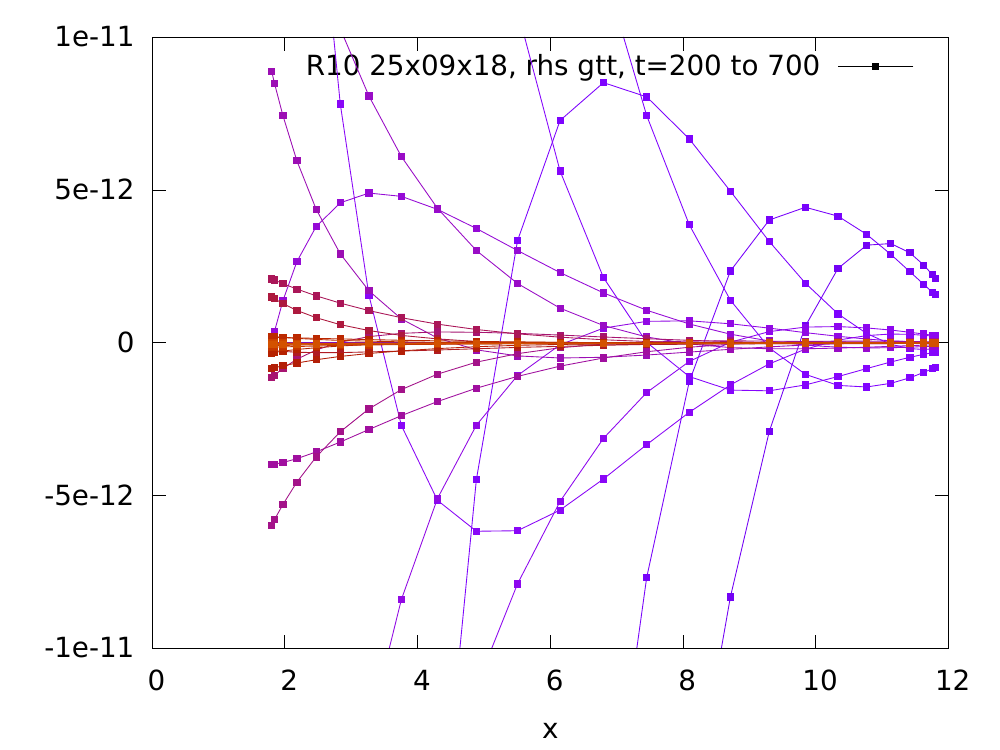}
  \caption{
    Single BH, grid R10.
    Some variables on the $x$-axis for different times.
    Top left: Variables $g_{tt}$, $g_{tx}$, and $g_{xx}$ at time $t=0$ and
    $t=1000$.  
    Top right: Time derivative of $g_{tt}$, $g_{tx}$, and $g_{xx}$ at $t=0$. 
    Bottom left and right: Time derivative of $g_{tt}$ during the
    evolution at two different scales. There is an oscillation in
    $g_{tt}$ that is largest for small $r$. The color coding indicates
    early times in dark, later times in brighter colors. The amplitude
    of the oscillation quickly decreases with time.
  }
  \label{r10_x}
\end{figure}

\begin{figure}[t]
  \centering
  \includegraphics[width=120mm]{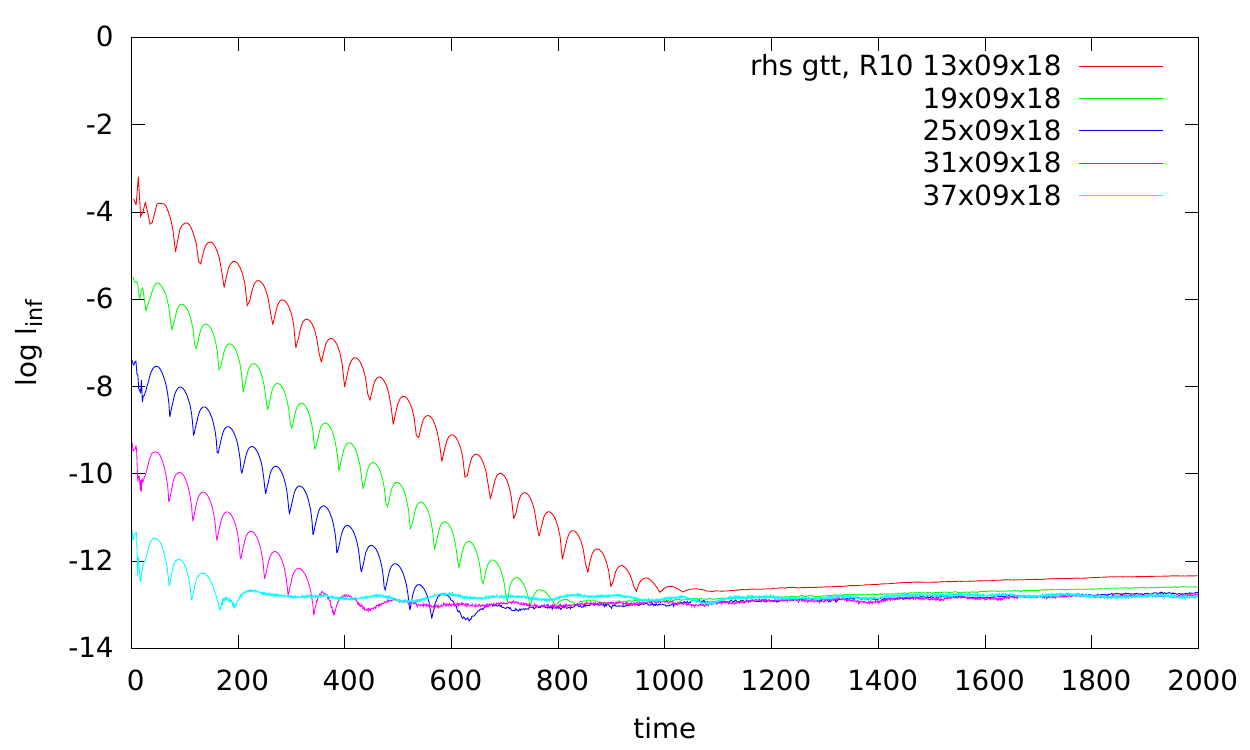}
  \caption{
    Single BH, grid R10. Shown is the logarithm of the infinity-norm of the
    right-hand-side of the evolution equation for the variable
    $g_{tt}$ versus time. 
    The number of grid points in the radial direction is varied 
    from $N_r=13$ to
    $37$ while keeping the angular resolution fixed at $N_\theta=9$
    and $N_\phi=18$.  
    The analytic initial data leads to a finite error that depends on
    the spatial resolution of the grid. Key feature of these runs is
    the exponential convergence with radial resolution, and
    that the system settles down in an approximately
    stationary state of the discretized equations. 
  }
  \label{r10_rconv}
\end{figure}
\begin{figure}[t]
  \centering
  \includegraphics[width=80mm]{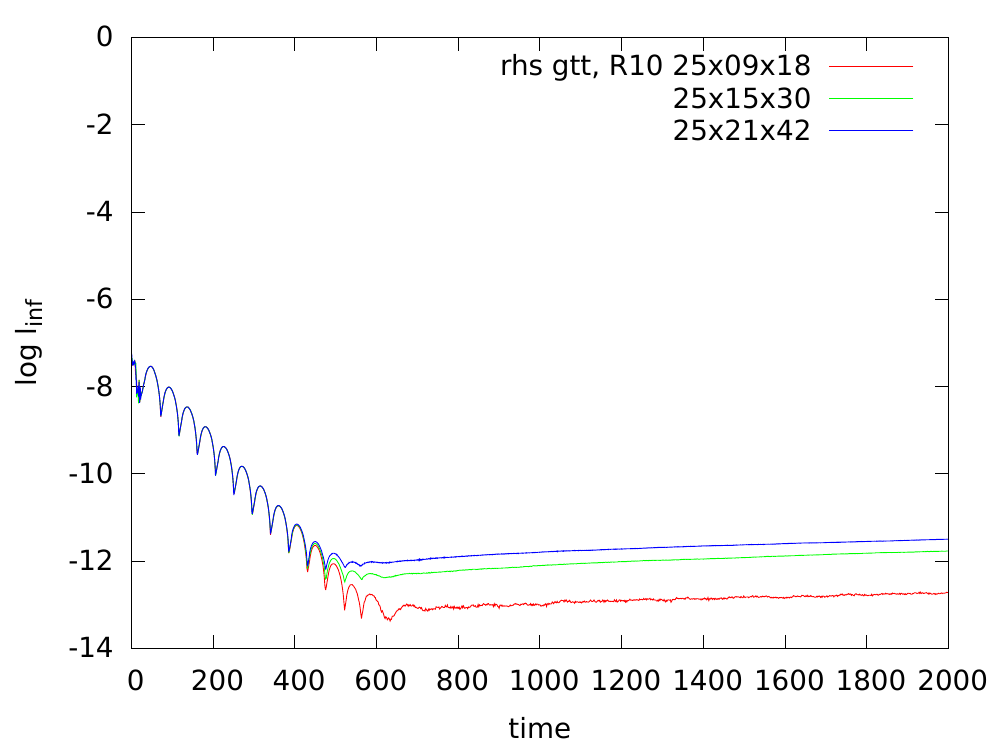}
  \includegraphics[width=80mm]{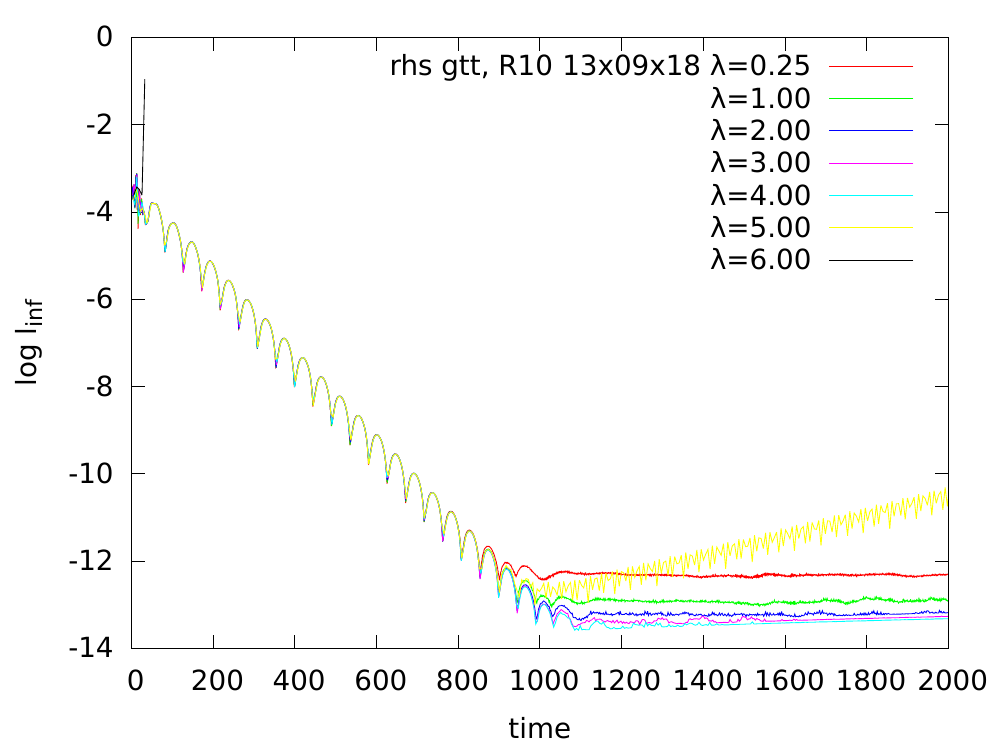}
  \caption{
    Single BH, rhs of $g_{tt}$.
    Left: Different angular resolutions at fixed radial resolution.
    The oscillations do not depend on angular resolution, but the
    round-off floor rises with resolution.
    Right:
    For the given grid, 
    varying the time step size, $\Delta t = \lambda \min(\Delta x)$,
    leads to stable runs for $\lambda \lesssim 4$ and to
    unstable runs for $\lambda \gtrsim 5$.  The stable runs settle
    down within $t=1000$ of evolution time. The oscillations in the rhs
    (both the period and amplitude) are independent of $\lambda$ down
    to $10^{-12}$, i.e.\ the discretization error associated with RK4
    is less than $10^{-12}$. The larger the number of time steps, the
    larger the error for the stationary regime beyond $1000M$, ranging
    from $10^{-12}$ for $\lambda = 0.25$ to $10^{-13}$ for
    $\lambda = 4.0$.  The run for $\lambda = 5.0$ is borderline
    unstable, with a comparatively slow exponential growth. 
    The run for $\lambda = 6.0$ fails within $50M$. 
  }
  \label{r10dtfa}
\end{figure}

In Fig.~\ref{r10_x}, we consider the evolution on a grid with
dimension $25\times9\times18$. 
Shown are the metric components 
$g_{tt}$, $g_{tx}$, and $g_{xx}$ and some of their time derivatives
on the $x$-axis at different times $t$. In the top left, we show the
initial data at $t=0$ and the data at $t=1000$. On this scale, 
no evolution is discernible, that is the lines for $t=1000$ fall on
top of the lines for $t=0$.
In the top right, we show the numerical right-hand-side (rhs) for the
variables, i.e. the numerical approximation to $\partial_t g_{tt}$,
$\partial_t g_{tx}$, and $\partial_t g_{xx}$. At $t=0$, they are
non-zero at around $10^{-8}$. This indicates that already for the
small grid with $25\times9\times18$ points the spectral method gives a
rather accurate approximation to the analytic solution, for which
the time derivatives vanish. 
In the bottom left and right of Fig.~\ref{r10_x}, the time evolution
of $\rgtt$ is shown for vertical scales of $10^{-8}$ and
$10^{-11}$. The color/gray-scale coding shows progressing time from
dark to lighter colors. There is an oscillation that is largest for
small $r$. 
For the black hole, many quantities
follow an approximate $1/r^p$ dependence, with gravity being strongest near the
inner boundary and falling off for large radii.
The amplitude of the oscillation decreases with
time.

In Fig.~\ref{r10_rconv}, we show how the oscillations are damped with
time and how the solution converges with resolution.
As a representative example for the time dependence of the system, we
consider a norm of the right-hand-side of $g_{tt}$.
We plot the logarithm of the infinity-norm as a function of time, i.e.\
$\log_{10}(\max|\rgtt|)(t)$,
where the maximum is computed on the innermost sphere of the grid where the
fields are strongest.
We vary the radial resolution from
$N_r=13$ to $37$ while keeping the angular resolution fixed at
$N_\theta=9$ and $N_\phi=18$.

Fig.~\ref{r10_rconv} shows that, as expected, the analytic initial
data leads to a finite error that depends on the spatial resolution of
the grid. Key feature of these runs is stability and convergence, and
that during the time evolution the system settles down in an
approximately stationary state of the discretized equations.  While
settling down, the system oscillates with a frequency and amplitude
that is independent of the resolution. The time dependence dies out
exponentially.
Exponential convergence with radial resolution is evident.  Round-off
error is reached around $10^{-12}$ to $10^{-13}$. In this simple case,
$N_r\approx40$ suffices to approximate the initial data and the
evolution at round-off accuracy.

In Fig.~\ref{r10dtfa}, on the left we examine the dependence of
$\rgtt$ on angular resolution. Although the initial data is 
spherically symmetric, since the numerical method is fully 3d
deviations from sphericity occur. The initial, damped oscillations do
not depend on the angular resolution. However, the level of the round-off
error increases when the number of grid points is increased.

\begin{table}[t]
\centering
\begin{tabular}{|c|cc|cc|c|}
\hline
$ N_r \times N_\theta \times N_\phi $ &
$ \Delta t_{stab} $ &
\footnotesize
$ \begin{array}{l} \Delta t_{unst} \\ - \Delta t_{stab} \end{array} $ &
$ \frac{ \Delta t_{stab} }{ \min(\Delta r) } $ &
$ \frac{ \Delta t_{stab} }{ \min(\rho\Delta\phi) } $ &
$ \lambda_{\max} = [\frac{ \Delta t_{stab} }{ \min(\Delta x) } ] $
\\\hline
$13\times09\times18$& 0.4881 & 0.0036  &  2.865 &  4.497  &   4.4 \\
$19\times09\times18$& 0.4489 & 0.0060  &  5.910 &  4.135  &   5.9 \\
$19\times15\times30$& 0.2325 & 0.0030  &  3.060 &  5.910  &   5.9 \\
$25\times09\times18$& 0.2651 & 0.0015  &  6.198 &  2.442  &   6.1 \\
$25\times15\times30$& 0.2397 & 0.0031  &  5.603 &  6.093  &   6.0 \\
$31\times09\times18$& 0.1734 & 0.0009  &  6.332 &  1.598  &   6.3 \\
$31\times15\times30$& 0.1734 & 0.0009  &  6.332 &  4.409  &   6.3 \\
\hline
\end{tabular}
\caption{
Single BH. Empirical time-step size for stable evolutions with RK4. 
The data is based on a bisection search bracketed by 
values of the time step $\Delta t$ for stable and unstable runs.
Runs are called stable if they do not fail within $t=10000$. 
The largest stable time step size found is denoted $\Delta t_{stab}$, while
$\Delta t_{unst}$ is the smallest time step found for an unstable run. 
The result can be related to the smallest grid spacing in space, which
depending on $N_r$ and $N_\phi=2N_\theta$ may be obtained for the points 
in the radial direction, with clustering due to the Chebyshev grid, or for 
the $\phi$-direction, with points clustering near the poles.
In this example, if $\min(\Delta r)$ is less than 
$\min(r\sin\theta\Delta \phi)$,
then $\Delta t_{stab}$ is independent of $N_\phi$, 
and the time step can be chosen up to roughly 6 times larger 
than the smallest grid spacing, $\lambda_{\max} \lesssim 6$. 
}
\label{tab:dtfac}
\end{table}

In Fig.~\ref{r10dtfa}, right panel, we vary the time step size looking for
the largest allowed time step giving a stable evolution.  For
stability of the time integration, the rule of thumb is that the
eigenvalues of the pseudospectral spatial operator have to lie in the
stability region of the method of line integrator, although in general
this is not a sufficient condition and the pseudospectra have to be
considered \cite{Tre00}. The argument about domains of dependence
leading to a Courant-Friedrich-Lewy condition $v\Delta t/\Delta x\leq
const$, where $v$ is the propagation speed, does not apply directly to
pseudospectral methods since the spatial stencil covers the entire
domain.
Here we only investigate stability by numerical experiment.
Tab.~\ref{tab:dtfac} shows the result of a numerical, iterative search
for the largest allowed time step $\Delta t$. We find that this is
directly related to the smallest spatial distance on the grid.
For the 3d spherical grid 
defined in (\ref{rk})--(\ref{phij}) with $N_\phi=2N_\theta$, it depends
on the number of grid points in the different directions whether
the clustering of points in the radial or in the $\phi$-direction is
more severe. We either have 
$\min(\Delta x) = \min(\Delta r) = r_1-r_0$, or 
$\min(\Delta x) = \min(2r\sin(\theta)\sin(\Delta\phi/2)) \approx
  r_0 \sin(\theta_0) (\phi_1-\phi_0)$. 
It turns out that the ``Courant factor'' defined by
\beq
  \lambda = \Delta t/\min(\Delta x),
\eeq
determines stability, i.e.\ we should choose 
a time step $\Delta t = \lambda \min(\Delta x)$ with 
$\lambda<\lambda_{\max}$. Based on Tab.~\ref{tab:dtfac}, 
$\lambda_{\max} \approx 6$ for grids with $N_r\geq19$, even though
the smallest $\Delta x$ may occur in different directions. 
As a default, we choose $\lambda = 4$ in standard runs.

\setlength{\unitlength}{1mm}
\begin{figure}[t]
  \centering
  \begin{picture}(120,65)
  \put(40,0){\includegraphics[width=80mm]{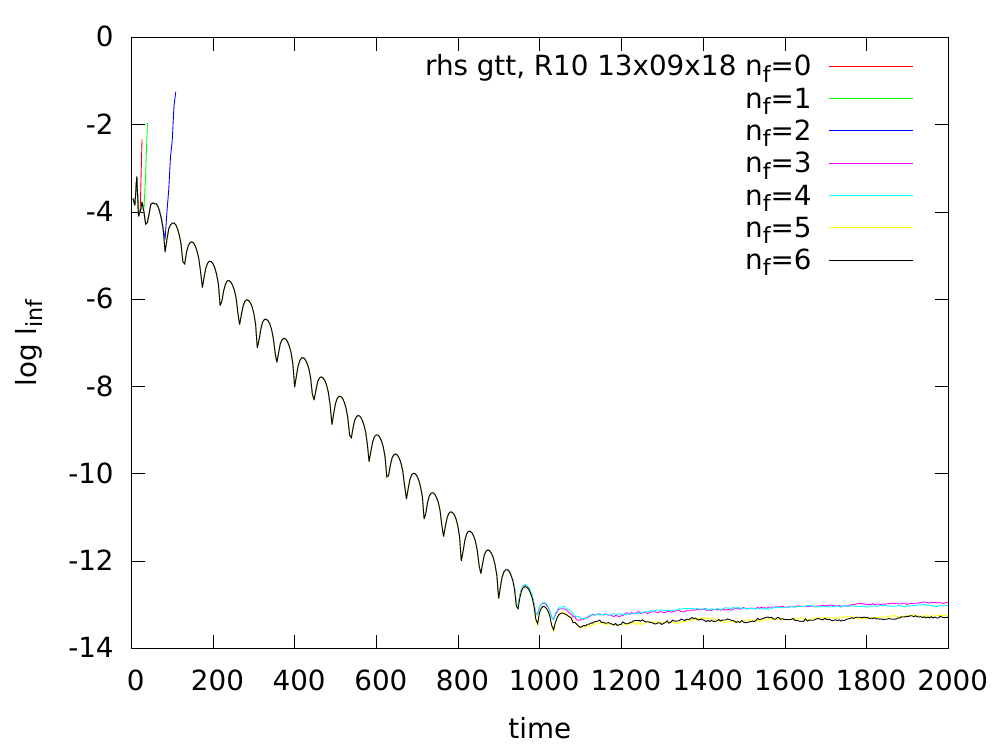}}
  \put( 0,20){\includegraphics[width=40mm]{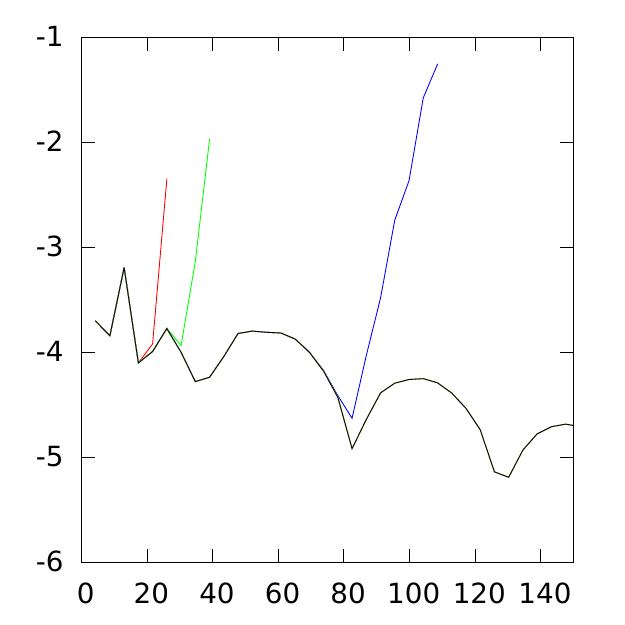}}
  \end{picture}
  \caption{
    Single BH, rhs of $g_{tt}$.
    Dependence on $n_f$, the number of spin-weighted
    spherical harmonics removed from the top for filtering. For
    $n_f=0,1,2$, the runs fail very quickly within $t=110$. For $n_f\geq3$, the
    runs appear stable, although for long runs, there are some cases where
    $n_f=3$ fails earlier than the others. In most cases we set $n_f=4$.
  }
  \label{r10_nf}
\end{figure}

In Fig.~\ref{r10_nf}, we show how the evolution
depends on the degree of tensor spherical harmonic filtering, with
$n_f$ indicating the number of modes that are set to zero in the
spherical harmonic projection.  For $n_f=0$, projection onto
tensor spherical harmonics is performed without additional filtering,
which nevertheless removes certain high-frequency components of the
double Fourier basis near the poles. For $n_f=0,1,2$, the runs become
unstable on a very short time-scale.
For $n_f\geq3$, the runs appear stable, although for long runs, there
are some cases where $n_f=3$ fails earlier than the others. Our
default choice is therefore $n_f=4$.
This behavior is consistent with the expectation that the tensor rank
of the fields determines the minimal $n_f$ required for stability. In
our example, the highest rank for the components in $u^\mu$ is 3,
which implies that spin-weights
$0,\ldots,\pm3$ occur in the tensor spherical harmonic decomposition.
For consistent filtering, it is not sufficient to only filter for
weights $<3$. Furthermore, we also require the derivatives $\partial_k
u^\mu$, which raises the rank to 4. Apparently, since we filter
the fields, even $n_f=3$ has a chance to work. In \cite{LinSchKid05},
the filter is applied to the right-hand-sides which are of rank 4, so
in that case $n_f\geq4$ may be strictly necessary.
Finally, we note that for the given experiment we do not seem to
require a filter based on, say, a 2/3 or 1/2 rule, due to the
non-linearity of the fields. Independent of the grid size, a constant
$n_f=4$ suffices to obtain rather long-term numerical stability.

\begin{figure}[t]
  \centering
  \includegraphics[width=80mm]{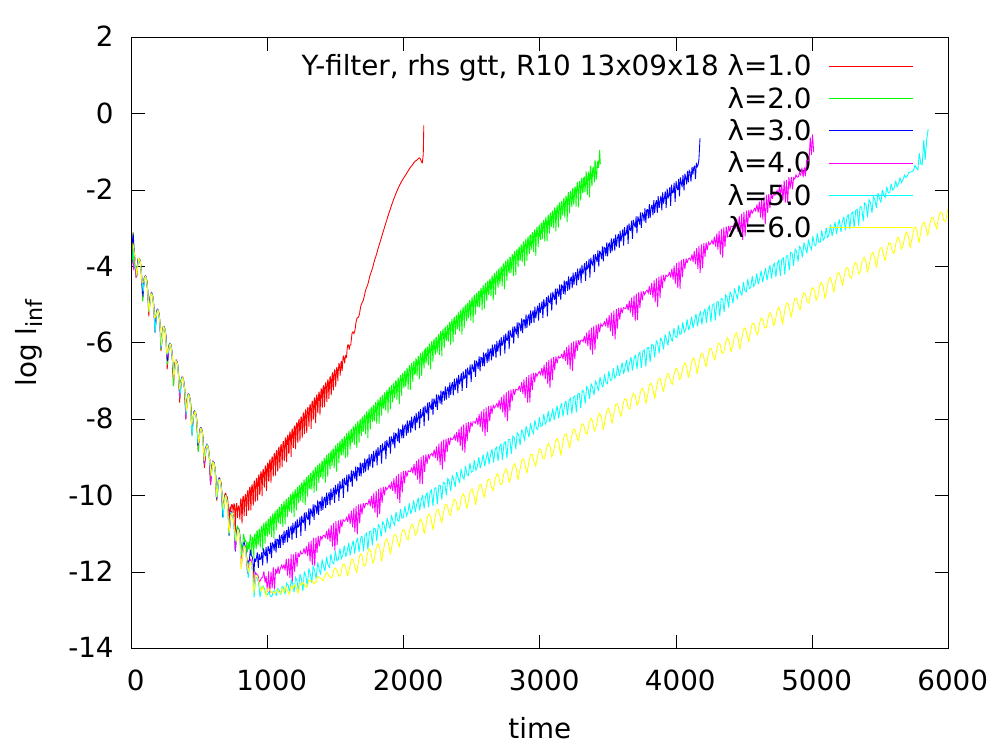}
  \includegraphics[width=80mm]{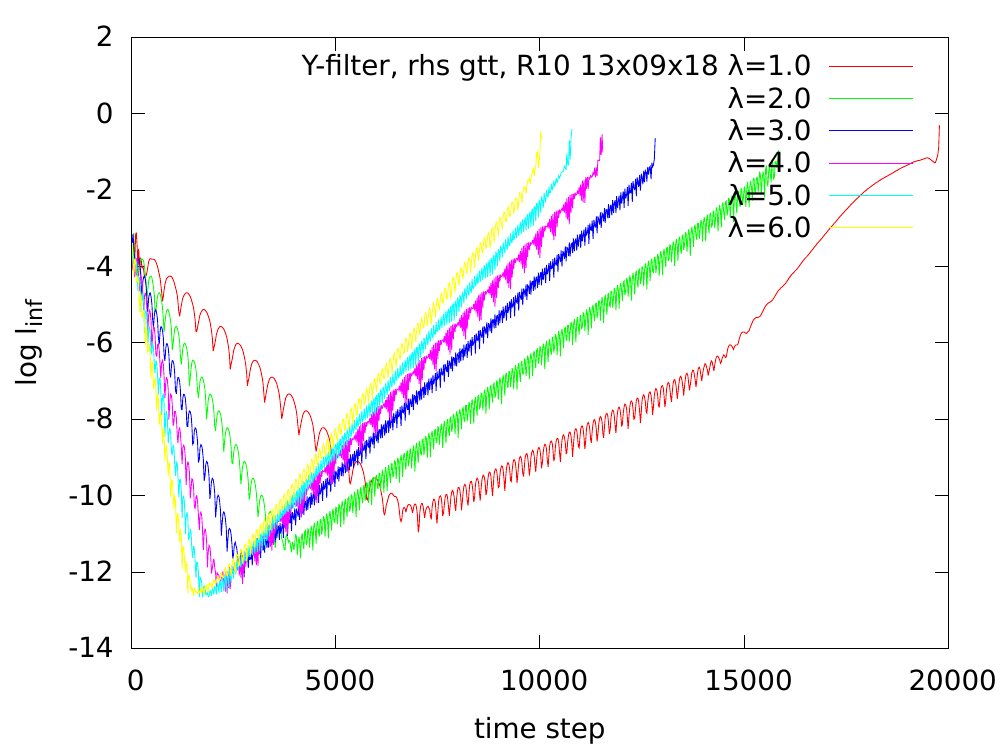}
  \includegraphics[width=80mm]{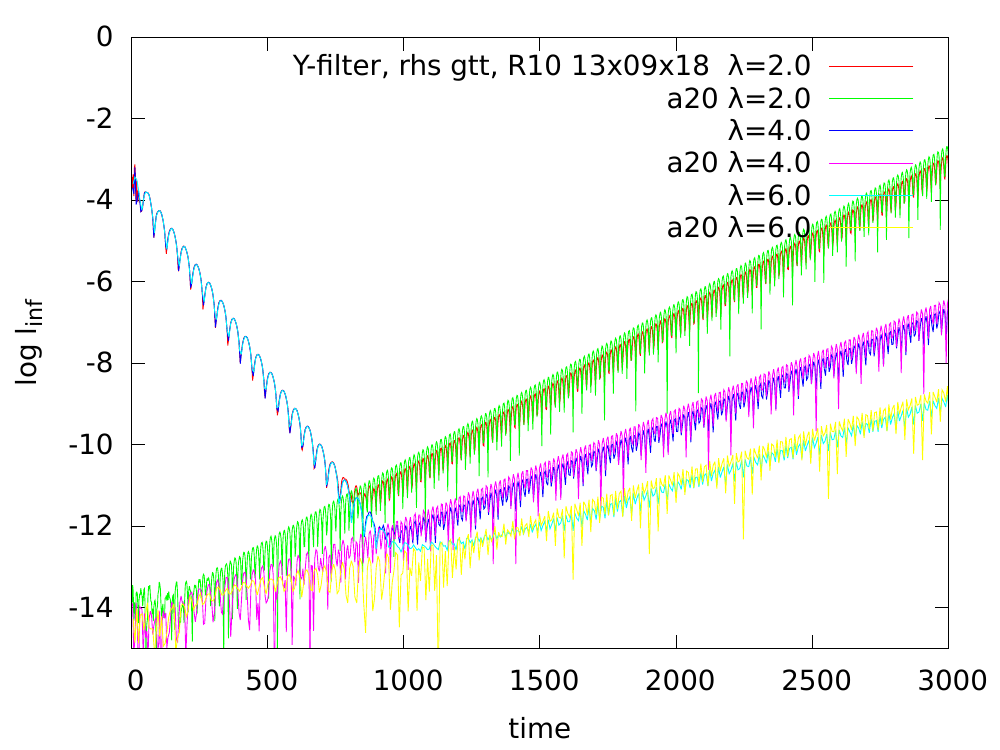}
  \caption{
    Single BH, rhs of $g_{tt}$. Runs with the scalar Y-filter fail within a
    time of a few thousand. Top: Runs for different Courant factors
    $\lambda=\Delta t/\Delta x_{\min}$ versus time (left) and versus
    number of time steps (right). Runs for larger $\lambda$ last
    longer. The exponential growth depends on the number of time
    steps, but it is not a simple proportionality. Bottom: The
    instability is directly related to the $a_{20}$ mode, which starts
    growing exponentially at about $t=0$ from around $10^{-14}$. 
  }
  \label{r10_f0}
\end{figure}

Runs without the spin-weighted Yn-filter are not as stable 
for the two alternatives that we tried. 
First, we consider the basic scalar Y-filter. This filter ignores the
tensor character of the components of $u^\mu$, but each component
is smooth and the approximation is spectrally convergent. 
In Fig.~\ref{r10_f0}, runs with the Y-filter and $n_f=4$ are seen to fail 
within a time of about 7000. Although
the initial damped oscillation is exactly that of the Yn-filter runs,
at about $t=1000$, there is exponential growth at a constant rate that
leads to the failure of the run. For $n_f\leq2$, the runs are much
shorter lived, while increasing $n_f$ to 4, 5, or 6 does not change
the picture, similar to the Yn-filter runs. This does not appear to
be a time-step instability due to choosing $\lambda$ too large, i.e.\
smaller $\lambda$ fail earlier. 
The exponential growth depends on $\lambda$, but it is not simply
proportional to the number of time steps, see the top right panel of
Fig.~\ref{r10_f0}. 
We also investigate the behavior of individual $a_{lm}$ and $b_{lm}$
modes in the expansion of $\rgtt$ in non-weighted spherical harmonics,
(\ref{realYexp}).
The bottom panel of Fig.~\ref{r10_f0} demonstrates
that the instability is directly related to the $a_{20}$ mode, which
starts growing at about $t=0$ from around $10^{-14}$, overtaking the
decay of the overall function at around $t=1000$. Other modes grow as
well, but we only show the largest mode. In other words, already at
early times there is a small error at round-off accuracy that is not
visible in $\rgtt$, which seeds an unstable, unphysical mode
that is not kept in check by the Y-filter.

\begin{figure}[t]
  \centering
  \includegraphics[width=80mm]{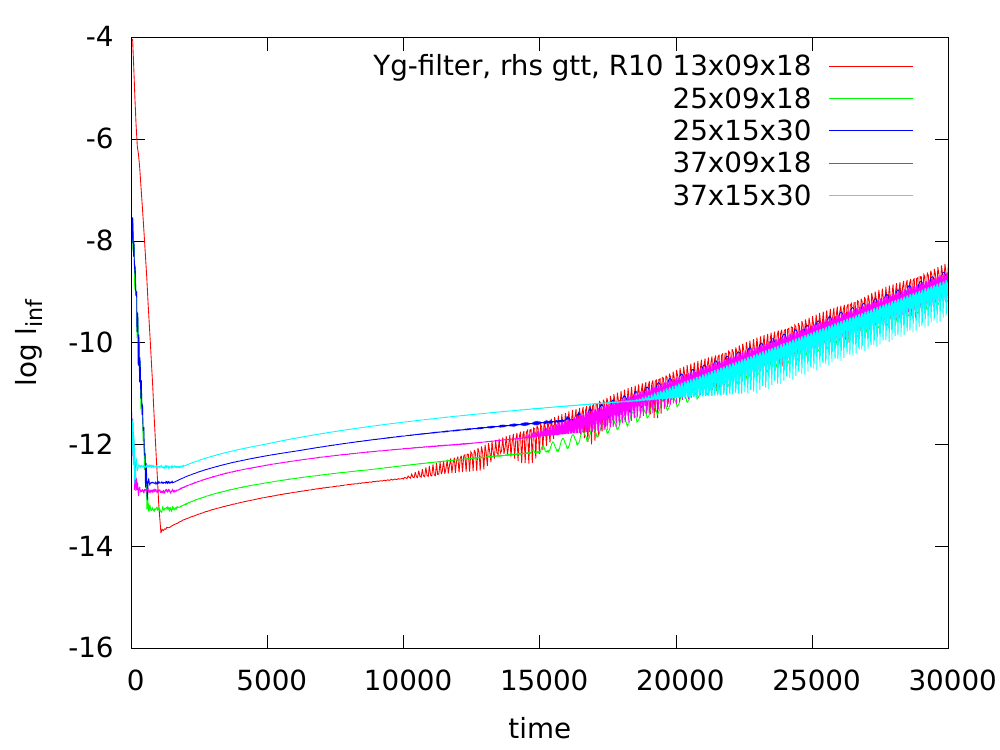}
  \includegraphics[width=80mm]{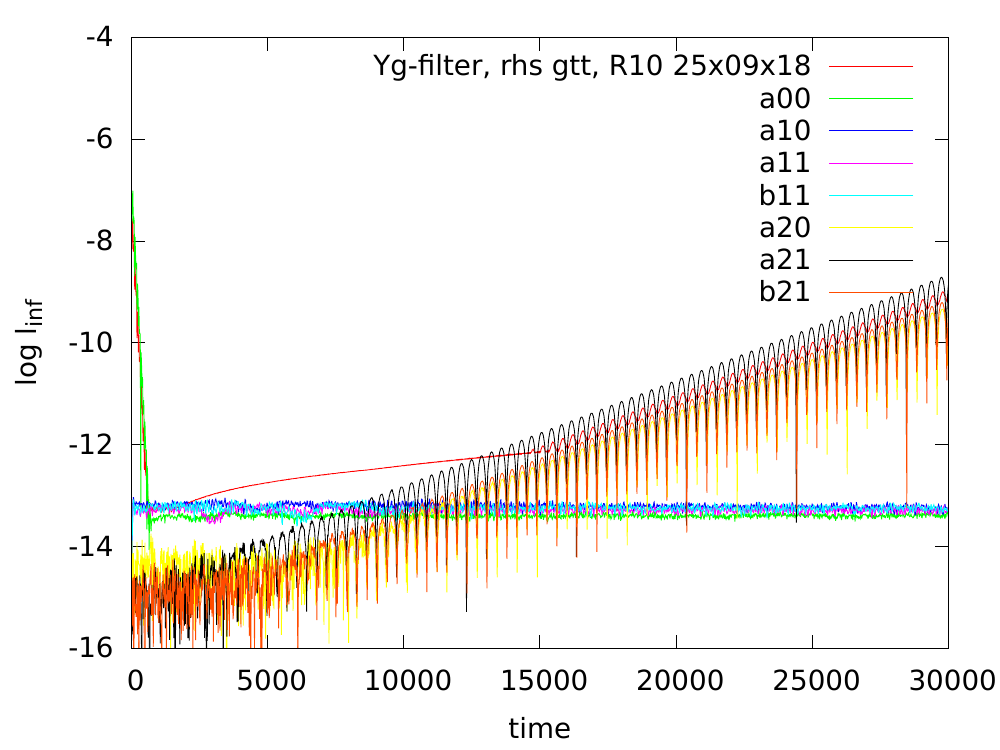}
  \caption{
    Single BH, rhs of $g_{tt}$.
    The graded Yg-filter allows simulations that last until
    about $t=70,000$. 
    On the left, we see how the run settles down exponentially
    by $t=1,000$, which is followed by a slow linear growth until about 
    $t=10,000$ to $20,000$,
    followed by exponential growth that eventually crashes
    the run. The rates of decay and growth are independent of radial
    resolution. 
    On the right, we show for a medium resolution how growth 
    in certain scalar spherical 
    harmonic modes starts dominating the behavior of the field.
  }
  \label{r10_f1}
\end{figure}

As an inbetween alternative to the Y- and Yn-filters, we consider the
graded Yg-filter with $n_f(\mu) = n_f - d(u^\mu)$, see the discussion
around (\ref{nfofmu}), which takes into account the shift between
tensors with a different number of spatial indices.
Fig.~\ref{r10_f1} shows results at five different resolutions.  The
runs settle down in the same manner as before within $t=1000$.  The
exponentially growing modes shown for the Y-filter in
Fig.~\ref{r10_f0} do not occur. After $t=1000$, there is some slow
linear growth. Computing $\rgtt$ for $t\leq10,000$ it may even appear
that there is no additional instability. However, there is another
type of exponential growth occuring at a slower rate than for the
Y-filter, which is also starting at round-off at early times. 
On the right in Fig.~\ref{r10_f1}, we show how various scalar harmonic
modes behave during the run at some particular resolution.  After the
initial phase, the modes $a_{00}$, $a_{10}$, $a_{11}$, and $b_{11}$
remain below $10^{-13}$. The main instability is visible in $a_{20}$,
$a_{21}$, and $b_{21}$. It appears to start at $t=0$ at around
$10^{-15}$, and then grows at a constant exponential rate. Notice how
the unstable modes overtake the regular feature at about $t=15,000$ in
the plot on the left.

\begin{figure}[t]
  \centering
  \includegraphics[width=150mm]{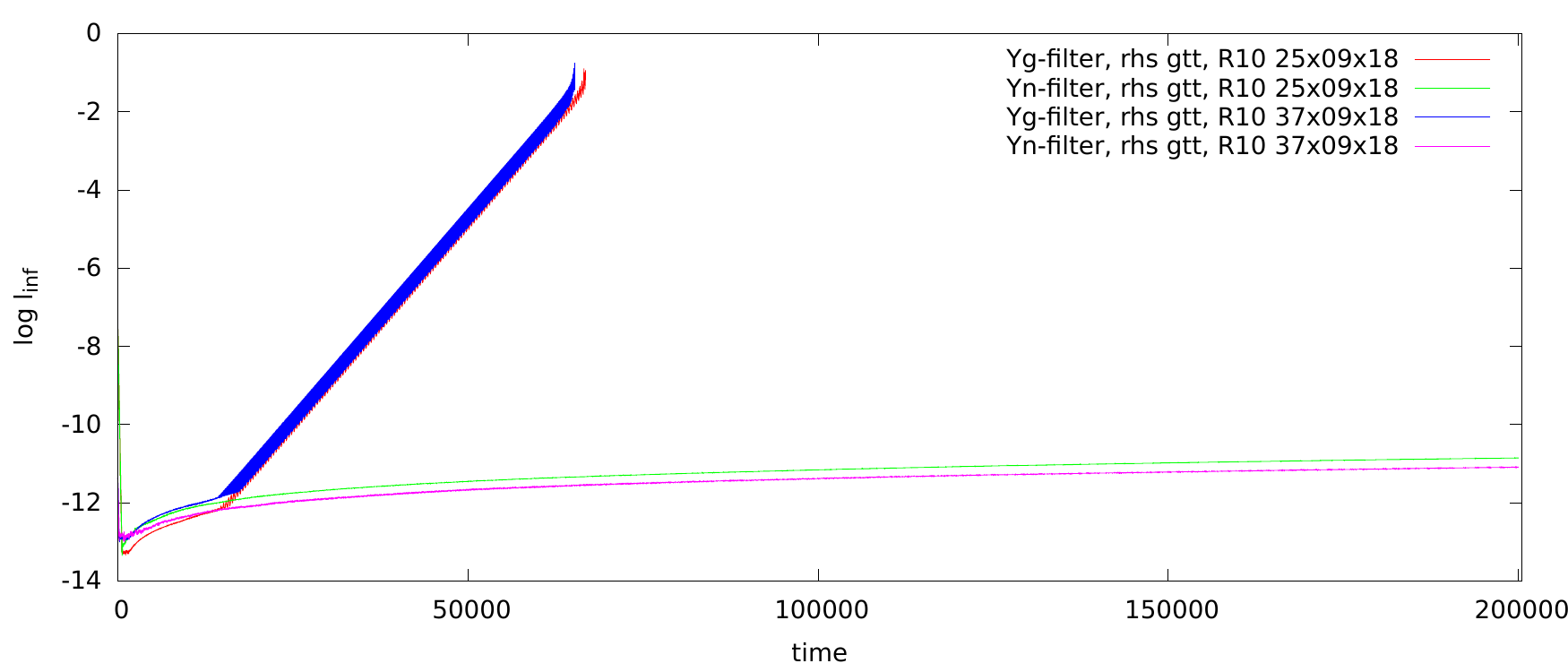}
  \includegraphics[width=75mm]{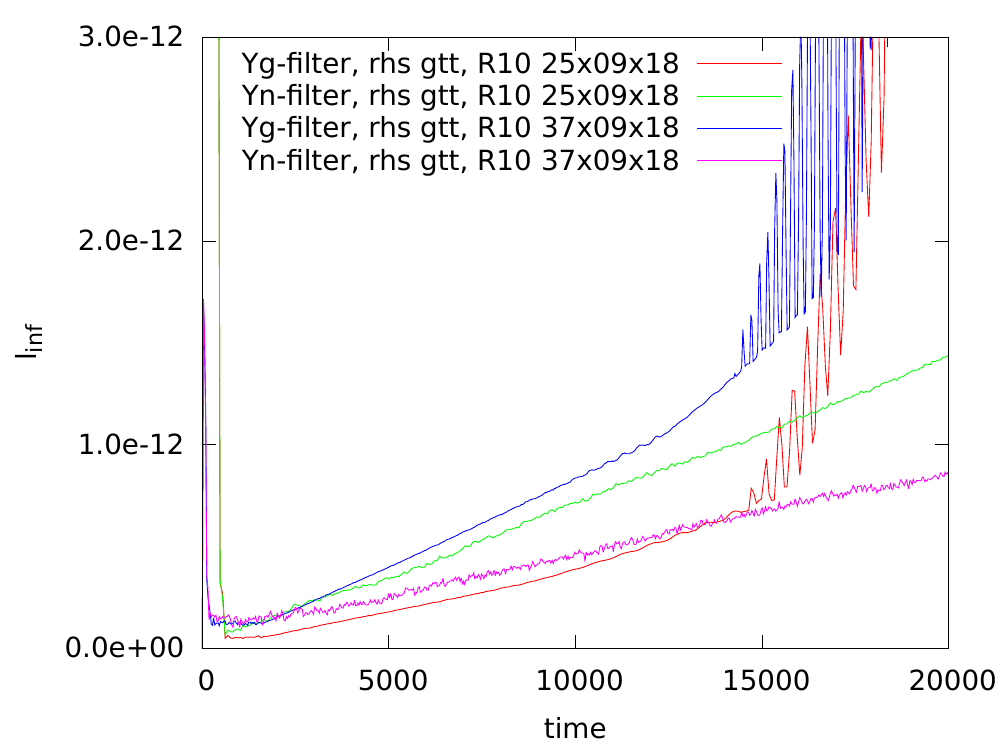}
  \includegraphics[width=75mm]{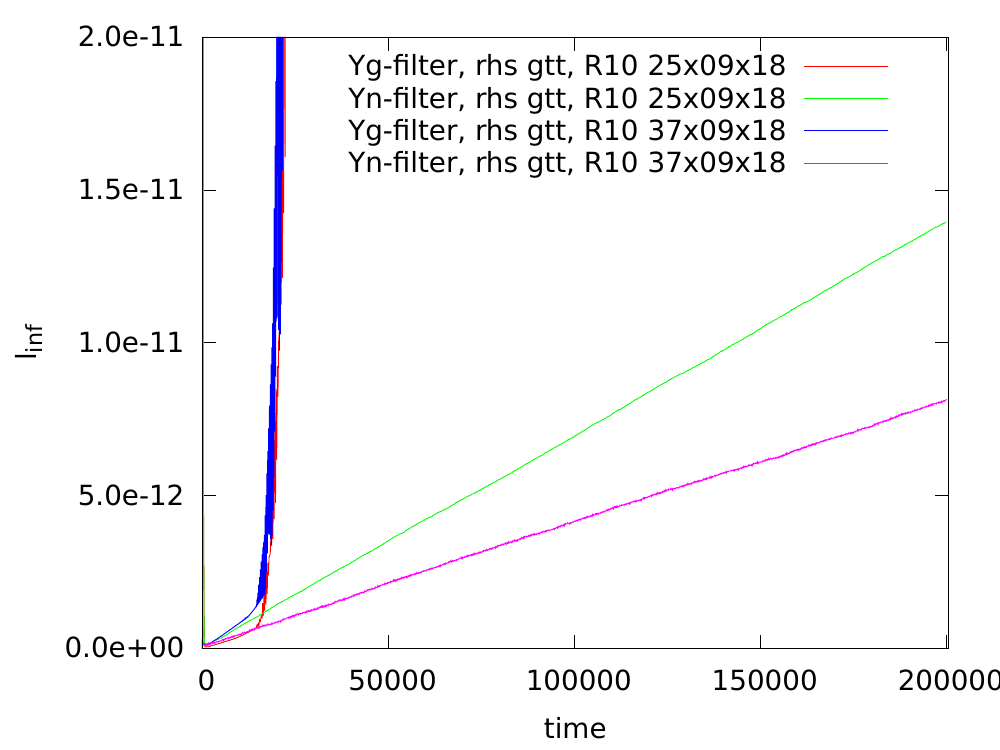}
  \caption{
    Single BH, rhs of $g_{tt}$.
    Long term stability, comparison between the graded Yg-filter and
    the tensor Yn-filter. Top: The Yg-filter runs fail around
    $t=70,000$, with an unstable mode visible at $t=15,000$ around
    $10^{-12}$. The Yn-filter runs last beyond $t=200,000$. 
    Bottom: For both filters there is a linearly growing mode. Its slope
    is roughly the same for both filters, and it is smaller for higher radial
    resolution.
  }
  \label{r10_long}
\end{figure}

\begin{figure}[t]
  \centering
  \includegraphics[width=80mm]{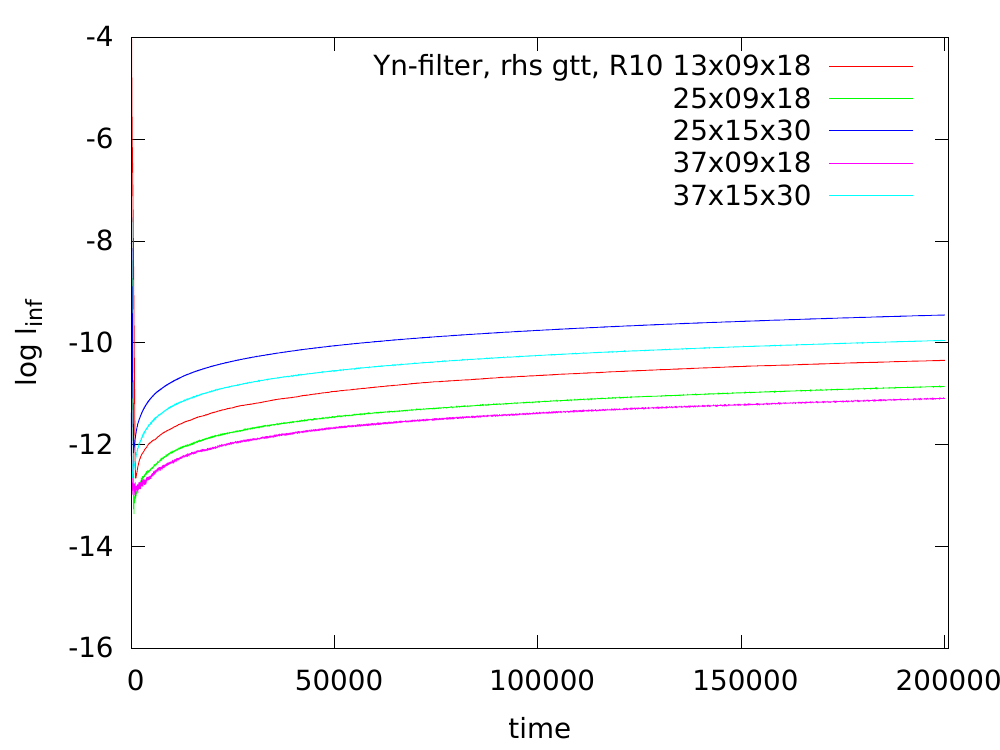}
  \includegraphics[width=80mm]{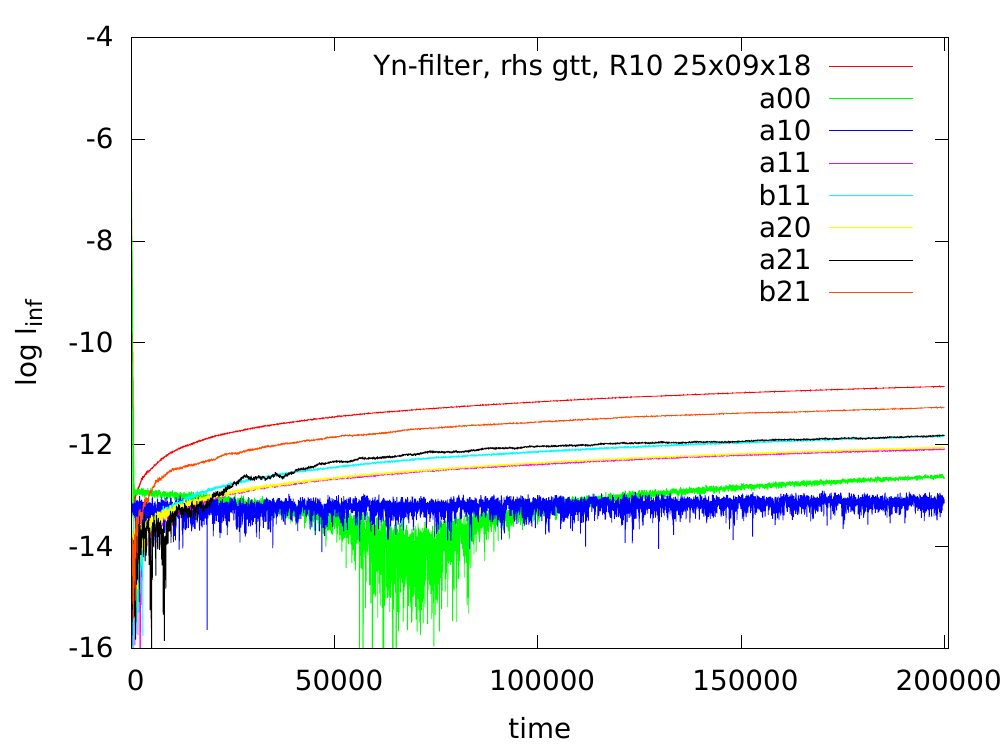}
  \caption{
    Single BH, rhs of $g_{tt}$. 
    Long-term behavior for the tensor
    Yn-filter. Shown are the same quantities as in
    Fig.~\protect\ref{r10_f1}. The runs last for at least
    $t=200,000$. A residual linear growth is visible, which is less
    than $10^{-10}$ per $\Delta t = 100,000$, depending on resolution.
  }
  \label{r10_f2}
\end{figure}

The Yn-filter cures both exponential modes that occur for the Y-filter
and the Yg-filter.  We compare the Yg-filter and the Yn-filter for
$t\leq200,000$ in Fig.~\ref{r10_long}. The Yn-filter runs do not
exhibit any exponential growth, although some linear growth remains,
see the bottom panels. The linear growth is roughly the same for both
filters, and it decreases with radial resolution. As far as the
tensor character of the fields is concerned, the Yn-filter with
$n_f\geq3$ should
remove all instabilities due to an inconsistent treatment of
tensors. However, other instabilities may well occur at a later time.
If so, they are not yet visible in the mode decomposition by
$t=200,000$, compare Fig.~\ref{r10_f2}. 

It may be worth recalling
that the target of state-of-the-art binary black hole simulations is
the last 10 or perhaps 20 orbits before merger, which corresponds
to $t\lesssim10,000$. If the limitations of the present example would carry
over, the method would comfortably satisfy the numerical stability
requirement. Note also that typical code tests only report evolution times
of 
up to $t=400$ in \cite{KidLinSch04},
$t=10,000$ in \cite{LinSchKid05},
or $t=5,000$ in \cite{Tic09}.
However, there is no reason that the simplest black hole test should
not yield unlimited stability. In fact, the main limitation of our
test case is that we ignore the available sophisticated outer boundary
conditions for the GHG system, e.g.\
\cite{LinSchKid05,RuiRinSar07,RinLinSch07,RinBucSch08}.  
As it turns out, simply increasing the radial dimension of the shell
by moving the outer boundary from $r_{\max}\approx 12$ to $22$ makes
the runs fail in a way that appears to signal a breakdown due to the
boundary condition. We leave the investigation of proper outer
boundary conditions to future work since our focus here is on the
construction of the CFF/Yn-method.

\begin{figure}[t]
  \centering
  \includegraphics[width=80mm]{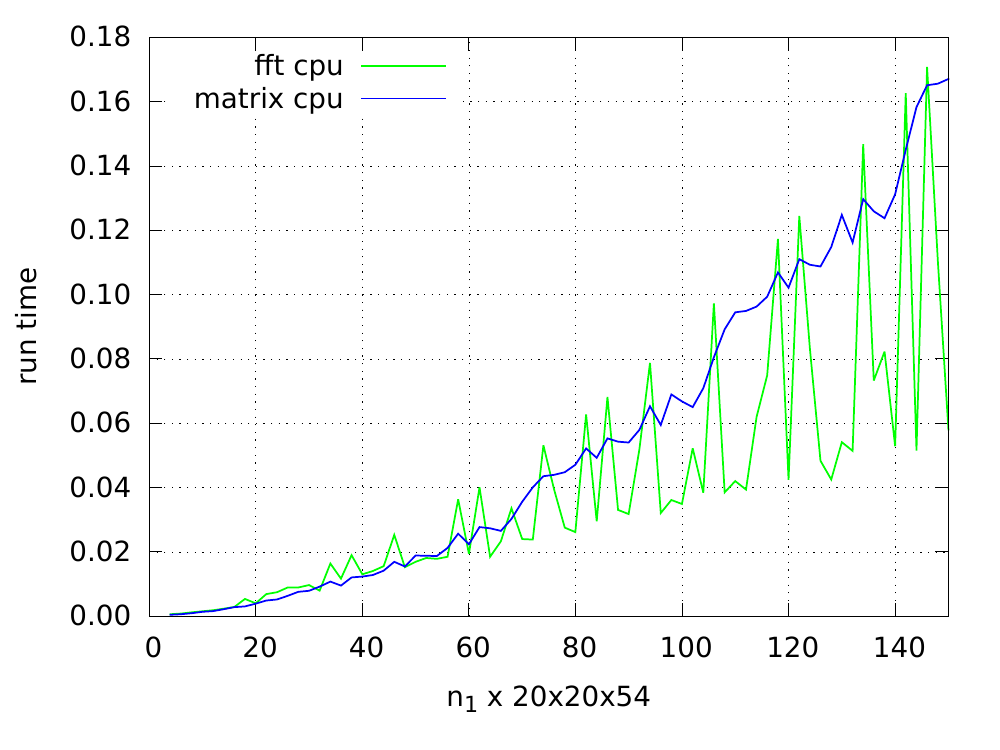}
  \includegraphics[width=80mm]{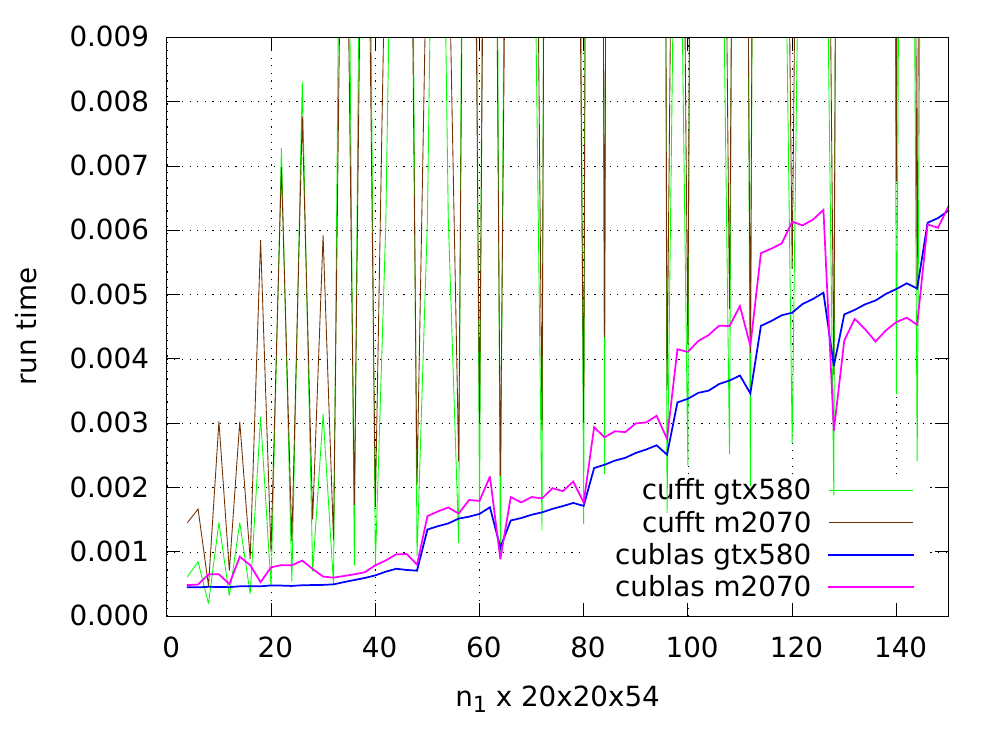}
  \caption{
    Example for the performance of matrix multiplication and the 
    fast Fourier transform relevant for pseudospectral differentiation
    on a CPU (left) and on two different GPUs (right). 
    Shown is the runtime versus the leading dimension $n_1$.
    We compare the multiplication of
    a $n_1\times n_1$ matrix and a $n_1\times(20\!\times\!20\!\times\!54)$
    matrix with the corresponding two Fourier transforms. 
    As is typical for FFT implementations, performance depends
    strongly on the problem size. 
    In this concrete example, on the CPU the matrix multiplication
    offers comparable performance for $n_1\lesssim70$, while on the
    GPU $n_1\lesssim100$ comparing with the $n_1$ for which the peak
    performance is compared. Considering all $n_1$, on average matrix
    multiplication is significantly faster than the FFT for the GPUs
    even beyond $n_1=150$.  
  }
  \label{mmfft}
\end{figure}

\subsection{Computational efficiency}
\label{efficiency}

The choice of matrix multiplication methods for both the spectral
derivatives and filters can be viewed as one of convenience, since it
simplifies the implementation of spectral methods on the sphere, in
particular for the tensor spherical harmonic filter.  However, a
priori it is not clear what the difference in performance is compared
to the fast Fourier transform. If there was a significant performance
penalty due to the $O(N^2)$ operations of matrix-vector
multiplications compared to the $O(N\log N)$ of FFTs, then we should
aim for a fast transform implementation (with the possible exception
of the Legendre transform). However, as argued above, the 3d physics
problem that we consider leads to transforms with $N\lesssim50$.  For such
small $N$, the matrix multiplication method can be even faster than
the FFT \cite{For98,Boy01}.

Fig.~\ref{mmfft} shows a representative benchmark for our specific
method. Part of the pseudospectral algorithm is the 1d transform of
several variables on a 3d grid, see (\ref{D1}) for the derivative
$\partial_x$ and (\ref{Fanalysis}) for the Fourier analysis in the
$\phi$-direction. In Fig.~\ref{mmfft}, we compare the run time as a
function of the leading dimension $n_1$ (assumed contiguous in memory)
for different implementations of the matrix multiplication of a
$n_1\times n_1$ matrix and a $n_1\times(20\!\times\!20\!\times\!54)$
matrix. For simplicity, we consider only this operation and do not 
include the computation of actual derivatives (the FFT method requires
a transformation in Fourier space) or the generalization to all three 
directions.
As example for CPU performance, we show results for a single core of a
i7-870 CPU using FFTW 3.2.2 for the Fourier transform and ATLAS 3.8.3
(sse3) for the matrix multiplication.
As example for GPU performance, we consider NVIDIA's GTX580 and M2070 
Fermi/Tesla cards running CUDA 3.2 versions of CUBLAS and CUFFT. 
Notice in Fig.~\ref{mmfft} that the Tesla card can outperform the GTX 
card only in special cases and only for the matrix multiplication.

As expected, the performance of the matrix multiplication scales
approximately like $N^2$, while FFT performance depends strongly on
the size of the transform (i.e.\ on the prime factors of $n_1$). We
vary $n_1$ in steps of two, $n_1=4,6,\ldots$. For the CPU, the matrix
multiplication offers comparable performance for $n_1\lesssim70$. For
the GPUs, comparable performance is obtained for $n_1\lesssim100$, but
only when comparing with the optimal values of $n_1$. Optimization for
arbitrary $n_1$ is currently not as even with CUFFT as with FFTW3.
Although it might be feasible to eventually restrict physics runs to
the available fast $n_1$-FFTs, we note that on the GPUs on average
matrix multiplication is significantly faster (by a factor of more
than 10) than the FFT method even beyond $n_1=150$.  We therefore
focus exclusively on the the matrix multipliation method in this work.

In Tab.~\ref{tab:gpumm}, we quote some results for the performance in
Gflop per second of the three matrix-matrix multiplications required
for the computation of 3d partial derivatives, see
Sec.~\ref{derivatives3d}. 
Optimization for the new Fermi chips was included in the transition
from CUDA 3.1 to 3.2. Certain small matrix multiplications
\cite{CUDA32bug} started
working on Fermi with CUDA 4.0rc.
The Tesla cards C2050 and M2070 have four
times the number of floating point units compared to the GTX series.
However, for the specific matrix sizes considered, a multiple of 64 is
required in the leading dimension to benefit from the additional FPUs.
For the small grids of the black hole example, we obtain around 50 to
100 Gflop/s. For somewhat larger grids the performance approaches 200
Gflop/s, reaching roughly 300 Gflop/s on the Tesla cards when the
leading dimension is 64. This is close to the maximal performance
reported for large square matrices in \cite{NatTomDon10}, on which the
current CUBLAS/Fermi optimization is based. The theoretical peak for
these Tesla GPUs is around 500 Gflop/s.

Tab.~\ref{tab:gpumm} does not include the four transpose operations of
the derivative calculation. In Tab.~\ref{tab:gpumemory}, we give
memory transfer rates in Gbyte/s for the transposes.  The Tesla card
had ECC memory activated.
Its transfer rate is only about half that of the GTX580 card. For the
medium grid sizes the GTX580 and the M2070 both achieve about 100
Gflop/s, so it appears that the M2070 compensates for the lower memory
speed with its larger number of floating point units.
We also compare with the bandwidth test for device-to-device copies
included in the SDK, which results in lower numbers than the peak
one-directional memory bandwidth of the cards (159, 192, and 150
Gflop/s, respectively).  The transposes are out-of-order copies that
on the Fermi cards reach half the speed of the direct device-to-device copies.
There probably is room for optimization of the transpose, but the
derivative calculation mostly depends on the speed of the matrix
multiplication.  In Tab.~\ref{tab:gpumemory}, we also quote Gbyte/s
for the matrix multiplication considered as a matrix to matrix copy
operation, and these numbers are lower by a factor of 4 to more than
10 than for the transposes.

\newcommand{\dddd}[4]
{$\begin{array}{r}#1\!\times\!#2\\\times#3\!\times\!#4\end{array}$}

\begin{table}[!t]
\centering
\begin{tabular}{|ll|cccc|}
\hline
MatMul
&
Gflop/s
&
\dddd{40}{20}{20}{54} &
\dddd{40}{40}{40}{54} &
\dddd{60}{60}{60}{54} &
\dddd{64}{64}{64}{54} 
\\\hline
GTX285 & 2.3    &   33 &  43 &  52 &  70
\\ 
GTX480 & 3.2rc1 &   -- & 101 &  88 & 163
\\ 
GTX580 & 3.2    &   -- & 117 & 104 & 192
\\ 
GTX580 & 4.0rc  &   77 &  99 &  94 & 177
\\ 
C2050  & 3.1    &   48 &  64 &  73 & 
\\ 
C2050  & 3.2rc1 &   -- & 108 & 103 & 284
\\ 
M2070  & 3.2    &   -- & 109 & 104 & 294
\\ 
\hline
\end{tabular}
\caption{
Performance of dgemm on GPUs/CUBLAS for the three matrix multiplications of
pseudospectral derivatives. Numbers are in Gflop per second. 
For the small grids of the black hole example, we obtain around
50 to 100 Gflop/s. For somewhat larger grids the performance approaches
200 Gflop/s, reaching 300 Gflop/s on the Tesla cards when the leading
dimension is 64. 
}
\label{tab:gpumm}
\end{table}

\begin{table}[!t]
\centering
\begin{tabular}{|ll|rc|cc|cc|cc|c|}
\hline
\multicolumn{2}{|c|}{
\begin{tabular}{c}
Transpose, \it MatMul
\\
Gbyte/s
\end{tabular}
}
&
\multicolumn{2}{c|}{\dddd{40}{20}{20}{54}} &
\multicolumn{2}{c|}{\dddd{40}{40}{40}{54}} &
\multicolumn{2}{c|}{\dddd{60}{60}{60}{54}} &
\multicolumn{2}{c|}{\dddd{64}{64}{64}{54}} &
bandwidthTest
\\\hline
GTX285 & 2.3    &   46 &\it5.0&  45 &\it4.3&  53 &\it3.5&   21 &\it4.4&  124
\\ 
GTX580 & 3.2    &   71 &--&   78 &\it12&   78 &\it6.9&   77 &\it12&  139
\\ 
M2070  & 3.2    &   34 &--&   40 &\it11&   38 &\it6.9&   42 &\it18&   85
\\ 
\hline
\end{tabular}
\caption{
Performance of the matrix transpose on GPUs/CUDA for the four
transposes used to implement the 3d pseudospectral derivatives. 
Numbers are in Gbyte/s. For comparison, the device-to-device copy
operation from the CUDA SDK (bandwidthTest) is given. The combined
transpose-matmul-transpose operation is dominated by the matrix
multiplication.
}
\label{tab:gpumemory}
\end{table}

\begin{table}[!t]
\centering
\begin{tabular}{|l|cc|rr|c|}
\hline
Grid
&
\begin{tabular}{c} GPU \\ Algebra \end{tabular}
&
\begin{tabular}{c} GPU \\ MatMul \end{tabular}
&
GPU
&
CPU
&
CPU/GPU
\\\hline
$25\times09\times18$ &  18\% &  80\% &   16$s$ &  165$s$ & 10.0 \\
$37\times15\times30$ &  25\% &  74\% &   49$s$ &  823$s$ & 16.7 \\
$49\times21\times42$ &  23\% &  77\% &  133$s$ & 2903$s$ & 21.8 \\
\hline
\end{tabular}
\caption{
Performance of the spectral evolution of a single black hole 
on a GPU (GTX580, CUDA 4.0rc) compared
to one core of a CPU (i7-870). For a given grid size, 1000 RK4
evolution steps are performed. The startup time is not counted. The
matrix multiplications of the 
derivatives and the filter amount to about 75\% of the runtime on the
GPU, while the remainder is mostly due to the (simple but memory
intensive) algebra of the
right-hand-side of the Einstein equations.
For the largest grid, the GPU implementation is about 20 times faster
than the single core CPU implementation. 
}
\label{tab:bhbench}
\end{table}

In Tab.~\ref{tab:bhbench}, we give benchmark results for the black hole
example. For a given grid size, 1000 RK4
evolution steps are performed. The startup time is not counted, but
all other parts of the calculation except input/output are included.
The CFF/Y-filter method is a combination of matrix multiplications and
transposes for the derivatives, and a collection of matrix
multiplications of varying size for the filter. One other costly part
of the computation is the algebra that is required in addition to the
derivatives on the right-hand-side of the evolution equation. For the
Einstein equations, this is a sizable problem with about 10000
floating point operations per grid point per RK4 update. Furthermore,
there are about 200 variables (50 fields and their derivatives) plus a
correspondingly large number of temporary variables used during the
calculation, which exceeds by far the number of registers of the Fermi
cards (less than 64 are available per thread). On a CPU, the algebra
amounted to less than 10\% of the overall calculation, but on the GPU
the matrix part is more efficiently parallelized in the current
implementation.

The bottom line is that the GPU implementation is 21.8 times faster
than the (single core) CPU implementation for the largest grid. 
For smaller grids the speedup is still on the order of 10 to 17. 
A quad-core implementation using a BLAS library led to a speed up of 2
to 3 for the part of the matrix multiplication. Extrapolated for the
complete algorithm for the largest grid, this would still leave a
speed-up of 7 to 11 on the GPU.

\section{Conclusion}
\label{conclusion}

We discussed the construction of a pseudospectral method for
non-linear, time-dependent tensor fields on a spherical shell. The
proposed CFF/Yn-method, i.e.\ a Chebyshev-double-Fourier basis
combined with a spin-weighted spherical harmonic filter, was
successfully implemented for the model problem of a single black hole.
We demonstrated that a matrix method for both the spectral derivatives
and the filter resulted both in analytic simplicity and also in ease
and efficiency of the numerical implementation. To this end, we
developed a matrix method for the discrete spin-weighted scalar
harmonic projection for arbitrary spin weight.  The parallelization of
the CFF/Yn-method was evaluated for a GPU implementation. Numerical
results for three different filter strategies were given, showing that
the simple Y-filter and the improved, graded Yg-filter lead to
instabilities, that however are cured by the tensor Yn-filter.

In future work we intend to report on the generalization of the
method to domain decompositions. For example, the method as described
here has already been successfully applied to multiple nested shells
for scalar waves. Spherical shells are one of the building blocks
for more general grids that are needed, for example, for two black
holes. 
A key feature to implement for general applications in numerical
relativity is the incorporation of appropriate outer boundary
conditions.
In our example, a single black hole could be easily simulated in the
device memory of one GPU. Spectral methods may be able to realize an
appealing local optimum in computational efficiency, if two black holes
can be simulated on a single graphics card due to the memory
efficiency of the spectral method.

The GPU implementation is promising, giving speed-ups on the order of
10 to 20 compared to a single core of a CPU. A next step in the
optimization would be to use parallel kernels (which recently has
become possible for CUDA) for the different small matrix operations in
the spherical harmonic filter.  Although we focused on CUDA, the
important step is the organisation of the algorithm in terms of matrix
operations, which should be equally helpful for other platforms. In
general, pseudospectral matrix methods would benefit most from
further optimization of a subclass of matrix multiplications that is
somewhat outside the mainstream, i.e.\ the product of small, square
matrices with large, highly non-square matrices.

\appendix
\section{Formulation of the Einstein evolution problem}
\label{formulation}

In this section we collect the equations for the generalized harmonic gauge
(GHG) system and the single black hole test case. 
The GHG system that we focus on here is the
version introduced in~\cite{LinSchKid05}, which is a first order in
time and first order in space reformulation of the Einstein equations. A
first order harmonic system of this type was first considered
in~\cite{Alv02}, although well-posedness and numerical stability
requires the modifications introduced by~\cite{LinSchKid05}, in particular
constraint damping~\cite{BonLedPal03,GunGarCal05}. 
The generalized harmonic gauge was
introduced in \cite{Fri85}.  It played a major role in the binary
black hole evolutions of \cite{Pre04,Pre05,Pre06}, which used a second
order harmonic formulation. 
See \cite{Gar01,SziPolRez06} for further applications.
We give a short synopsis for the test case of a single black hole,
which nevertheless is quite complicated since the stability tests are
performed for the full Einstein equations in 3d.  Although much of the
material is contained or at least implicit in \cite{LinSchKid05}, it
should be helpful to readers not familiar with numerical relativity to
spell out some of the details.
The notation is adapted to the 3+1 decomposition of \cite{Yor79}.
\subsection{Einstein equations in generalized harmonic gauge}
\label{equations}

The goal is to solve the Einstein equations of classical general
relativity in vacuum, 
\beq 
  R_{ab}(g, \partial g, \partial^2 g) = 0, 
\eeq 
for the 4-metric $g_{ab}$.
The Ricci tensor $R_{ab}$ is constructed from first and second
derivatives of the metric. 
The construction of the GHG system starts with the observation that
the Ricci tensor can be written as
\beq
	R_{ab} = -\frac{1}{2} g^{cd} \partial_c\partial_d g_{ab}
                 + \partial_{(a} \Gamma_{b)} - g^{cd} \Gamma_{cab} \Gamma_d
                 + g^{cd} g^{ef}(\partial_e g_{ac} \partial_f g_{bd}
                                 - \Gamma_{ace} \Gamma_{bdf}),
\label{RabGamma}
\eeq
where we have introduced the Christoffel symbol of the metric and 
one of its contractions,
\bea
\Gamma_{cab} &=& \frac{1}{2} 
           (\partial_a g_{bc} + \partial_b g_{ac} - \partial_c g_{ab}),
\quad
\Gamma_c = g^{ab} \Gamma_{cab}.
\label{Gamma}
\eea
Note that in (\ref{RabGamma}) the second derivatives of the metric are
conveniently separated into $g^{cd} \partial_c\partial_d g_{ab}$ and
$\partial_{(a} \Gamma_{b)}$. The first term represents a standard wave
operator, while the second does not.

We can choose harmonic coordinate functions $x_a$ for which $\Gamma_a
= - \nabla^b\nabla_b x_a = 0$. In this gauge the principal part of the
Ricci tensor consists only of the wave operator, leading to a symmetric
hyperbolic system. Generalized harmonic
coordinates satisfy 
$
  \Gamma_a = - H_a
$
for some given gauge source functions $H_a$ that may depend on the
coordinates and the metric, but not on the derivative of the metric.
Since $H_a$ does not contribute to the principal part, we again arrive
at a second order symmetric hyperbolic system.
The GHG is based on a modified Einstein equation, where the
coordinates are incorporated through the constraint function $\calC_a
= H_a + \Gamma_a$, see in particular \cite{BonLedPal03} on the Z4
system. This suggests a constraint damping scheme which is essential
for the stability of the GHG system~\cite{GunGarCal05}. 

In order to discuss the GHG system as a Cauchy problem, we assume that
the coordinates naturally split into time and space, $x_a=(t,x_i)$.
The spacetime normal to the hypersurfaces of constant time $t$ is
given by $n_a= - \alpha \nabla_a t$, with the lapse function $\alpha$
chosen such that $n_an^a=-1$. The time-flow vector field $t^a=(1,0^i)$
is given by $t^a=\alpha n^a + \beta^a$, where the shift vector
$\beta^a$ is tangential to the hypersurface, $n_a\beta^a=0$. We have
\beq
	n_a = (- \alpha, 0_i),
\quad
	n^a = g^{ab} n_b = (\frac{1}{\alpha},-\frac{\beta^i}{\alpha}).
\label{normal}
\eeq
The 4-metric $g_{ab}$ induces a 3-metric $\gamma_{ij}$ in the
hypersurface, and determines lapse and shift,
\beq
	\gamma_{ij} = g_{ij},
\quad
        \beta_j = g_{tj}, 
\quad
	\beta^i = \gamma^{ij} \beta_j,
\quad
	\alpha = \sqrt{\beta_i\beta^i - g_{tt}}.
\label{abgfromgab}
\eeq
The inverse 4-metric is denoted by
$g^{ab}$, and the inverse 3-metric is denoted by $\gamma^{ij}$.
Notice that $g_{ij}=\gamma_{ij}$, but 
$g^{ij} = \gamma^{ij} - \beta^i\beta^j/\alpha^2$.
Raising and lowering indices for 4d indices is done with the 4-metric,
and for 3d indices the 3-metric is used.

A first order version of the GHG system can be obtained
straightforwardly by introducing new variables for the first
derivatives of the metric, 
\beq
	\myPhi_{iab} = \partial_i g_{ab},
\quad	\myPi_{ab} 
	= - \frac{1}{\alpha} (\partial_t g_{ab} - \beta^i\partial_i g_{ab}),
\label{defPiPhi}
\eeq
where the time derivative is in direction of the hypersurface normal,
$\myPi_{ab} = - n^c \partial_c g_{ab}$.  The resulting first order
system was first discussed in~\cite{Alv02}.  The modifications of
\cite{LinSchKid05} for stability involve constants $\gamma_0$,
$\gamma_1$, $\gamma_2$, and $\gamma_3$. Choose $\gamma_3=
\gamma_1\gamma_2$ to obtain symmetric hyperbolicity for all $\gamma_1$
and $\gamma_2$.  Choose $\gamma_1 = -1$ to avoid certain shocks. Less
clear is the choice of $\gamma_0$, which controls the Gundlach-type
constraint damping involving the gauge constraint $\calC_a = \Gamma_a
+ H_a$, and the choice of $\gamma_2$, which appears as a coefficient
of the constraint $\calC_{iab} = \partial_i g_{ab} - \myPhi_{iab}$ due
to the introduction of first order variables.  We choose $\gamma_0=1$
and $\gamma_2=1$, which is reported to lead to stable evolutions in
standard numerical experiments \cite{LinSchKid05}.

The GHG system in first order form including the modifications for stability 
takes the form
\beq
	\partial_t u^\mu = - {A^{k\mu}}_\nu \partial_k u^\nu + S^\mu,
\label{dotu}
\eeq
where $u^\mu = \{g_{ab}, \myPi_{ab}, \myPhi_{iab}\}$ is the vector of
variables, and where ${A^{k\mu}}_\nu$ and $S^\mu$ depend on
$u^\mu$ but not its derivatives. 
Written out explicitly,
\bea
	\partial_t g_{ab} &=& - \alpha \myPi_{ab} + \beta^i \myPhi_{iab},
\label{gdot}
\\
	\partial_t \myPhi_{iab}
	&=& 
	\beta^k \partial_k \myPhi_{iab}
	- \alpha \partial_i \myPi_{ab}
	+ \alpha \partial_i g_{ab}
	+ \frac{1}{2} \alpha n^c n^d \myPhi_{icd} \myPi_{ab}
	+ \alpha \gamma^{jk} n^c \myPhi_{ijc} \myPhi_{kab} 
	- \alpha \myPhi_{iab},
\label{Phidot}
\\
	\partial_t \myPi_{ab} 
	&=&
	\beta^k\partial_k \myPi_{ab} 
        - \alpha \gamma^{ik} \partial_k \myPhi_{iab} 
	- \beta^k \partial_k g_{ab}
	+ 2 \alpha g^{cd}(\gamma^{ij}\myPhi_{iac}\myPhi_{jbd} 
	- \myPi_{ac}\myPi_{bd}
	- g^{ef} \Gamma_{ace} \Gamma_{bdf})
\nonumber
\\
	&& - 2\alpha \nabla_{(a} H_{b)} 
	- \frac{1}{2} \alpha n^c n^d \myPi_{cd} \myPi_{ab}
	- \alpha n^c \myPi_{ci} \gamma^{ij} \myPhi_{jab}
\nonumber
\\ &&
	+ \alpha (2 \delta^c_{(a} n_{b)} - g_{ab}n^c) (H_c+\Gamma_c)
	+ \beta^i\myPhi_{iab}.
\label{Pidot}
\eea
These equations assume that $H_a$ is given, for example through an additional
evolution equation. The complete system involves a state vector
$u^\mu = \{g_{ab}, \myPi_{ab}, \myPhi_{iab}, H_a\}$ or even
$u^\mu = \{g_{ab}, \myPi_{ab}, \myPhi_{iab},$ $H_a, \partial_t H_a \}$, 
if the evolution of $H_a$ is specified by an equation that is second order
in time. Since $g_{ab}$ and the other variables are symmetric in $a$
and $b$, there are 50, 54, or 58 variables in
$u^\mu$, respectively.

Since constraint conservation is non-trivial, once a set of
independent variables has been chosen we have to strictly distinguish
between dependent and independent variables.  For example, the
variable $\myPhi_{iab}$ and the first spatial derivative of the variable
$g_{ab}$ are treated separately, since they are only equal if the
corresponding constraint $\calC_{iab}=0$ is satisfied.
We collect the relations needed to compute the dependent quantities
appearing in (\ref{gdot})--(\ref{Pidot}) from the $u^\mu$. 
Quantities obtained from the 3+1 split of $g_{ab}$ are $n^a$,
$\alpha$, $\beta^i$, and $\gamma_{ij}$, see (\ref{normal})--(\ref{abgfromgab}).
The inverse metrics $g^{ab}$ and $\gamma^{ij}$ are computed as
inverses from the component matrices. 
The ``covariant'' derivative of
$H_a$ is defined as
$
	\nabla_aH_b = \partial_a H_b - g^{cd} \Gamma_{cab} H_d.
$
The Christoffel symbols $\Gamma_{abc}$ and $\Gamma_a$ are computed from
(\ref{Gamma}) using the following expressions for $\partial_c g_{ab}$ in
terms of undifferentiated dynamical variables,
\beq
	\partial_i g_{ab} = \myPhi_{iab},
\quad
\partial_t g_{ab} = - \alpha \myPi_{ab} + \beta^i \myPhi_{iab}, \eeq
compare (\ref{defPiPhi}) and (\ref{gdot}). All other partial
derivatives appearing in (\ref{gdot})--(\ref{Pidot}) including
$\partial_a H_b$ are computed directly (e.g.\ numerically) as
derivatives of the $u^\mu$, as required for the formulation (\ref{dotu}).

\subsection{Boundary conditions}

The boundary conditions imposed on the evolution system play a crucial
role in achieving well-posedness and numerical stability. The topic
has received a lot of attention, and there is a number of usually
rather complicated boundary conditions. These are often more
complicated than the evolution equations themselves since e.g.\ for
constraint conservation at the boundary the time evolution of the
constraints is needed, which requires higher than second order
derivatives of the metric.
For the investigation of boundary conditions, 
we summarize the
characteristic eigenvalue problem of the GHG system following
\cite{LinSchKid05,KidLinSch04}.
Consider a normalized spatial vector $s^i$ with
$s_is^i=\gamma_{ij}s^is^j=1$.
When considering 2d boundaries within constant time hypersurfaces, the vector
$s^i$ is the outward pointing unit normal to the boundary. The
eigenvalue problem associated with (\ref{dotu}) in direction $s^i$ is 
\beq
   {e^{\hat\alpha}}_\mu s_k {A^{k\mu}}_\nu 
   = v_{(\hat\alpha)} {e^{\hat\alpha}}_\nu,
\eeq
where the characteristic matrix is $s_k {A^{k\mu}}_\nu$, the left
eigenvectors are denoted by ${e^{\hat\alpha}}_\mu$, and the eigenvalues
by $v_{(\hat\alpha)}$. The index $\hat\alpha$ labels eigenvalues and
eigenvectors (and is not summed over on the right-hand-side). 
${e^{\hat\alpha}}_\mu$ depends on $s^i$, which typically depends on space and
time due to the normalization with respect to $\gamma_{ij}$.

We derive some explicit expressions for the first order GHG system
(\ref{gdot})--(\ref{Pidot}), where we suppress 
the ten components in the symmetric tensor indices.
With $b=s_k\beta^k$,
\beq
	u^\mu = \left(\begin{array}{c}
                g \\ \myPi \\ \myPhi_{i}
                \end{array}\right),
\qquad
s_k {A^{k\mu}}_\nu	
= \left(\begin{array}{ccc}
	0  &  0  &  0 \\
        b  & -b  &  \alpha s^i \\
        -\alpha s_i  &  \alpha s_i  &  -b
    \end{array}\right).
\eeq
Mathematica finds, considering the $(sA)^T$ right eigenvector equation,
transposing the result to obtain the left eigenvector matrix with the
eigenvectors written in the rows, scaling the eigenvectors for
convenience, and ordering the eigenvalues and eigenvectors to
correspond more closely to \cite{LinSchKid05},
\beq
v_{(\hat\alpha)} = 
\left(\begin{array}{c}
0
\\ 
+\alpha -b
\\ 
-\alpha -b
\\
-b 
\\ 
-b  
\end{array}\right),
\qquad
{e^{\hat\alpha}}_\mu = 
\left(\begin{array}{ccccc}
1 & 0 & 0 & 0 & 0 
\\
-1 & 1 & s^1 & s^2 & s^3
\\
-1 & 1 & -s^1 & -s^2 & -s^3
\\
0 & 0 & -s_3 & 0 & s_1
\\
0 & 0 & -s_2 & s_1 & 0
\end{array}\right).
\label{characteristics}
\eeq
This representation assumes that $s_1\neq0$. 
The two eigenvectors for eigenvalue $-b$ are orthogonal to $s_i$.
Alternatively, \cite{LinSchKid05} write $u^{\hat 2}_i = {P^k}_i
\myPhi_k$ for three fields obtained by orthogonal projection, all with
eigenspeed $-b$. 
We introduce the projection onto directions tangential to the boundary
and orthogonal to the boundary normal, ${P^k}_i=\delta^k_i - s^k s_i$.

The standard way to impose boundary conditions for symmetric
hyperbolic systems is to impose conditions on the characteristic
fields. To this end,
we split the partial derivatives in (\ref{dotu}) with
$\delta^k_i={P^k}_i+ s^k s_i$ and project (\ref{dotu}) onto
eigenvectors, resulting in
\beq
	{e^{\hat\alpha}}_\mu \partial_t u^\mu
        =
        - v_{(\hat\alpha)} {e^{\hat\alpha}}_\mu s^k \partial_k u^\mu 
        - {e^{\hat\alpha}}_\mu {P^k}_i {A^{i\mu}}_\nu \partial_k u^\nu
        + {e^{\hat\alpha}}_\mu S^\mu.
\label{edeltu}
\eeq 
In other words, we obtain an advection equation in the direction of
$s^i$ with characteristic speeds given by the eigenvalues,
plus terms involving derivatives tangential to the boundary. 
Equation (\ref{edeltu}) allows us to specify boundary conditions
that distinguishes between incoming and outgoing modes according to
$v_{(\hat\alpha)}<0$ and $v_{(\hat\alpha)}>0$, respectively.

Here we focus on the simplest condition that
is successful for the single black hole test case.
For the case of a Schwarzschild black hole,
\cite{LinSchKid05} reports that 
freezing the incoming characteristic fields, 
\beq
        \left. {e^{\hat\alpha}}_\mu \partial_t u^\mu
        \right|_{{boundary}} = 0
\quad \mbox{for $v_{(\hat\alpha)} < 0$,}
\label{freezeincoming}
\eeq
gives stable
evolutions. Therefore the most basic stability test does not involve
the complicated constraint characteristics. 
Since we evolve the $u^\mu$ and not the characteristic fields,
a boundary condition on the characteristic fields
cannot be implemented directly. Instead, 
we transform (\ref{dotu}) with
${e^{\hat\alpha}}_\mu$, set some of the time derivatives to zero
according to (\ref{freezeincoming}), and transform back with the
inverse of ${e^{\hat\alpha}}_\mu$. %
This procedure can be combined into a transformation by a single
matrix $E^{-1}ZE$, where $Z$ is a diagonal matrix with 0 on the
diagonal if $v_{(\hat\alpha)} < 0$ and 1 on the diagonal if
$v_{(\hat\alpha)} \geq 0$.

\subsection{Test case of a single, spherically symmetric and static black hole}
\label{initialdata}

As test case we consider the Schwarzschild spacetime, which describes
a single, spherically symmetric and static black hole
\cite{MisThoWhe73a}.  We write the Schwarzschild metric in Kerr-Schild
form,
\beq
  g_{ab} = \eta_{ab} + f l_a l_b, 
  \quad f = \frac{2M}{r}, \quad l_a = (1, \frac{x_i}{r}),
\label{KSmetric}
\eeq
where $\eta_{ab}$ is the Minkowski metric in coordinates $(t,x_i)$,
$r = (\delta^{ij} x_i x_j)^\frac{1}{2} = (x^2+y^2+z^2)^\frac{1}{2}$,
and $M$ is the mass of the black hole. The horizon is located at
$r=2M$.
A specific feature of the Kerr-Schild form is that the vector $l^a$ is
null ($l_al^a=0$) with respect to both $\eta_{ab}$ and $g_{ab}$.  We
have chosen to scale $l_a$ such that $l_il^i=\eta^{ij}l_il_j=1$, so
$l^i=\delta^{ij}l_j$ is the normalized radial vector with respect to
the Euclidean 3-metric.

The metric (\ref{KSmetric}) solves the Einstein equations,
and all coordinate time derivatives vanish, $\partial_tg_{ab} = 0$.
We have chosen geometrical units, $G=c=1$. Furthermore,
we set the black hole mass to one, $M = 1$,
so all quantities including length and time are dimensionless.
The numerically experiment consists of posing initial data based on
(\ref{KSmetric}) for $t=0$, and to study the numerical evolution of
this data. 

Initial data $u^\mu=\{g_{ab},\myPi_{ab},\myPhi_{iab}\}$ at $t=0$ is
computed from $g_{ab}(t,x_i)$, (\ref{KSmetric}), using the definition
of the first order variables, (\ref{defPiPhi}).  We compute the
spatial derivatives in (\ref{defPiPhi}) numerically from (\ref{KSmetric}).
In general, $\myPi_{ab}$ requires in addition the time derivative of the
metric. However, for our example, $\partial_t g_{ab} = 0$. 

We perform the evolution in the generalized harmonic gauge, where the
gauge source function is initialized based on the Kerr-Schild metric
(\ref{KSmetric}), which is a non-constant function of the $x_i$,
and which is left constant during the evolution, 
\beq 
  H_a(t=0) = -\Gamma_a(t=0), 
  \quad 
  \partial_t H_a = 0.  
\label{gauge}
\eeq
In \cite{LinSchKid05}, the gauge condition for this test case is not stated
explicitly, but based on \cite{RuiRinSar07} we assume that (\ref{gauge}) was
used, since it is equivalent to initializing $H_a$ with the condition
that $\partial_t\alpha = 0$ and $\partial_t\beta^i=0$. 
We note in passing that in a different formulation for the same type
of Kerr-Schild black hole it was found that the gauge functions
$\alpha$ and $\beta^i$ have to be allowed to evolve in order for a
numerically stationary solution to be found \cite{AlcBru00}.
In the present case, the gauge source $H_a$ is static, but
nevertheless lapse and shift can evolve.

We conclude with a comment on the characteristic speeds of the GHG
system for the Kerr-Schild metric.
In terms of 3+1 variables, the Kerr-Schild metric becomes
\beq
  \gamma_{ij} = \delta_{ij} + f l_i l_j,
\quad
  \alpha = \frac{1}{(1+f)^{1/2}}, 
\quad
  \beta^i = l^i \frac{f}{1 + f}.
\eeq
The outward pointing normal
$s_i$ to a boundary surface of constant $r$ is proportional to $l_i$,
but normalized with respect to $\gamma_{ij}$. 
Since 
$\gamma^{ij} = \delta^{ij} - \frac{f}{1+f}l^il^j$, we have 
$s_i = l_i/\sqrt{\gamma^{ij} l_il_j} = l_i \sqrt{1+f}$, and 
$b=s_i\beta^i=f/\sqrt{1+f}=f\alpha$. 
According to (\ref{characteristics}), the characteristic speeds
$v_{(\hat\alpha)}$ assume values 0, $\pm\alpha-b=(\pm1-f)\alpha$, and
$-b=-f\alpha$.  For $r\rightarrow\infty$, $b\rightarrow0$ and
$\alpha\rightarrow1$. Asymptotically for large distances, the speeds
are therefore 0 and $\pm1$. Only the mode with speed $\alpha-b$ can be
positive for the given data. 
It vanishes at the horizon at $r=2$, and $\alpha-b>0$ for
$r>2$.  At the horizon and actually for all $r<2$, all speeds are
$\leq0$, so at and below the horizon there are no incoming modes and
no extra boundary condition is required.
Since the eigenvalues are linked to the time-stepping stability, we
note that in the numerical example $r_{\min}=1.80$, and the eigenvalues range
from $-1.45$ to some value less than $+1$ at the outer boundary.
For $r_{\min}=2.00$, the fastest eigenspeed is $-\sqrt{2}=-1.41$, for
$r_{\min}=1.50$ it is $-1.53$.

At the outer boundary, the modes for
$\hat\alpha=2,3,4$ with $v_{(\hat\alpha)}=-\alpha-b, -b, -b$ are
incoming. These are the modes that we freeze for the simplistic
boundary condition (\ref{freezeincoming}). In this special case,
the combined transformation $E^{-1} Z E$ becomes
\bea
B(\partial_t g_{ab}) = \partial_t g_{ab},
\quad
B(\partial_t \myPi_{ab}) &=& \frac{1}{2} 
  (+ \partial_t g_{ab} + \partial_t
  \myPi_{ab} + s^k \partial_t \myPhi_{kab} ),
\\
B(\partial_t \myPhi_{iab}) &=& \frac{1}{2}
  s_i (- \partial_t g_{ab} + \partial_t
  \myPi_{ab} + s^k \partial_t \myPhi_{kab} ),
\eea
where $B$ denotes the boundary values for the right-hand-sides
determined by freezing the incoming modes.

\section{Examples for spin-weighted spherical harmonics}

We write the spin-weighted spherical harmonics as $Y^n_{lm} = \hat
P^n_{lm}(\cos\theta) e^{im\phi}$. The $\hat P^n_{lm}$ are normalized
Wigner $d$-functions, which in this context could also be called
normalized spin-weighted associated Legendre polynomials. The first
few $\hat P^n_{lm}$ for $n>=0$ are: 
\beq 
\begin{array}{lll}
\hat{P}^0_{0,0}=\frac{1}{2 \sqrt{\pi }} & \hat{P}^1_{0,0}=0 & \hat{P}^2_{0,0}=0 \\[3mm]
 \hat{P}^0_{1,-1}=\frac{1}{2} \sqrt{\frac{3}{2 \pi }} \sin (\theta ) & \hat{P}^1_{1,-1}=-\frac{1}{2}
   \sqrt{\frac{3}{\pi }} \cos ^2\left(\frac{\theta }{2}\right) & \hat{P}^2_{1,-1}=0 \\
 \hat{P}^0_{1,0}=\frac{1}{2} \sqrt{\frac{3}{\pi }} \cos (\theta ) & \hat{P}^1_{1,0}=\frac{1}{2}
   \sqrt{\frac{3}{2 \pi }} \sin (\theta ) & \hat{P}^2_{1,0}=0 \\
 \hat{P}^0_{1,1}=-\frac{1}{2} \sqrt{\frac{3}{2 \pi }} \sin (\theta ) & \hat{P}^1_{1,1}=-\frac{1}{2}
   \sqrt{\frac{3}{\pi }} \sin ^2\left(\frac{\theta }{2}\right) & \hat{P}^2_{1,1}=0 \\[3mm]
 \hat{P}^0_{2,-2}=\frac{1}{4} \sqrt{\frac{15}{2 \pi }} \sin ^2(\theta ) & \hat{P}^1_{2,-2}=-\sqrt{\frac{5}{\pi
   }} \cos ^3\left(\frac{\theta }{2}\right) \sin \left(\frac{\theta }{2}\right) & \hat{P}^2_{2,-2}=\frac{1}{2}
   \sqrt{\frac{5}{\pi }} \cos ^4\left(\frac{\theta }{2}\right) \\
 \hat{P}^0_{2,-1}=\frac{1}{2} \sqrt{\frac{15}{2 \pi }} \cos (\theta ) \sin (\theta ) &
   \hat{P}^1_{2,-1}=-\frac{1}{2} \sqrt{\frac{5}{\pi }} \cos ^2\left(\frac{\theta }{2}\right) (2 \cos (\theta
   )-1) & \hat{P}^2_{2,-1}=-\sqrt{\frac{5}{\pi }} \cos ^3\left(\frac{\theta }{2}\right) \sin \left(\frac{\theta
   }{2}\right) \\
 \hat{P}^0_{2,0}=\frac{1}{8} \sqrt{\frac{5}{\pi }} (3 \cos (2 \theta )+1) & \hat{P}^1_{2,0}=\frac{1}{2}
   \sqrt{\frac{15}{2 \pi }} \cos (\theta ) \sin (\theta ) & \hat{P}^2_{2,0}=\frac{1}{4} \sqrt{\frac{15}{2 \pi }}
   \sin ^2(\theta ) \\
 \hat{P}^0_{2,1}=-\frac{1}{2} \sqrt{\frac{15}{2 \pi }} \cos (\theta ) \sin (\theta ) &
   \hat{P}^1_{2,1}=-\frac{1}{2} \sqrt{\frac{5}{\pi }} (2 \cos (\theta )+1) \sin ^2\left(\frac{\theta }{2}\right)
   & \hat{P}^2_{2,1}=-\sqrt{\frac{5}{\pi }} \cos \left(\frac{\theta }{2}\right) \sin ^3\left(\frac{\theta
   }{2}\right) \\
 \hat{P}^0_{2,2}=\frac{1}{4} \sqrt{\frac{15}{2 \pi }} \sin ^2(\theta ) & \hat{P}^1_{2,2}=\sqrt{\frac{5}{\pi }}
   \cos \left(\frac{\theta }{2}\right) \sin ^3\left(\frac{\theta }{2}\right) & \hat{P}^2_{2,2}=\frac{1}{2}
   \sqrt{\frac{5}{\pi }} \sin ^4\left(\frac{\theta }{2}\right)
\end{array}
\eeq
For $n<0$, we have $\nP^{n}_{lm} = (-1)^{n+m} \nP^{-n}_{l,-m}$. For
example, $\nP^{-1}_{lm} = - (-1)^m \nP^{+1}_{l,-m}$. If $n=0$, we can
avoid the computation for $m<0$, but in general we need either the
$m<0$ or $n<0$ computation in order to obtain the other terms
through a simple sign flip.

There is no orthogonality for different spin weights, i.e.\ 
$(Y^{n'}_{l'm'},Y^n_{lm})$ can be non-zero even though $n'\neq n$. 
For example,
\beq
(Y^{-1}_{11},Y^1_{11}) = \frac{1}{2}, \quad
(Y^{0}_{11},Y^1_{11}) = \frac{3 \pi }{8 \sqrt{2}}, \quad
(Y^{1}_{11},Y^1_{11}) = 1, \quad
(Y^{1}_{11},Y^2_{21}) = \frac{5 \sqrt{15} \pi }{64}.
\eeq
As an example for expanding a vector component, consider the
unit vector in the $x$-direction,
$v^i=\delta^i_1$. One of its tetrad components is $v^im_i= m_x$,
which has the expansions
\bea
  m_x &=& \frac{1}{\sqrt{2}}(\cos\theta \cos\phi -i \sin\phi)
       = \sqrt{\frac{2 \pi }{3}} \left(Y^1_{1,1}-Y^1_{1,-1}\right)
\label{mxfinite}
\\
  &=&
\frac{1}{128} \sqrt{3} \pi ^{3/2} \left(
16 (Y^0_{1,-1}+Y^0_{1,1})+
4 \sqrt{5} (Y^0_{2,-1}-Y^0_{2,1})+
\sqrt{14} (Y^0_{3,-1}+Y^0_{3,1}) + \ldots
\right).
\eea
$m_x$ is a two term linear combination of the $Y^1_{lm}$, but
an infinite series in terms of the $Y^0_{lm}$ (and also the $Y^{-1}_{lm}$). 
The vector component
$\theta_x=\frac{1}{\sqrt{2}}(m_x + \barm_x) = \cos\theta \cos\phi$
does not have a finite series representation for any specific $n$,
since it is the linear combination of two vectors with different spin
weight. Not only do we obtain infinite series in specific cases, but
they typically converge only slowly because terms like $\cos\theta
\cos\phi$ are not continuous as functions on the sphere. However,
the spin-weighted harmonics are defined in such a way that the specific
discontinuities introduced by the complex tetrad vectors are exactly
resolved, e.g.\ as in (\ref{mxfinite}).

\section*{Acknowledgments}

It is a pleasure to thank David Hilditch, Andreas Weyhausen,
Gerhard Zumbusch,
and also Marcus Ansorg and Wolfgang Tichy
for discussions.
This work was supported in part by 
DFG grant SFB/Transregio~7 ``Gravitational Wave Astronomy''.


%
\bibliographystyle{model1-num-names}
\bibliography{refs,refsextra}

\end{document}